\documentclass[%
reprint,
nofootinbib,
 amsmath,amssymb,
 aps,
]{revtex4-1}

\usepackage{graphicx}
\usepackage{dcolumn}
\usepackage{bm}
\usepackage{longtable}
\usepackage{color}

\newcommand{\diff}[3][]{\frac{\partial^{#1}{#2}}{\partial{#3}^{#1}}}
\newcommand{\intdif}[3]{\int_{#1}^{#2}\!\!\!\textnormal{d}{#3}}
\newcommand{\expo}[1]{\textnormal{e}^{#1}}

\begin{document}


\title{Probing nuclear observables via primordial nucleosynthesis}

\author{Ulf-G.~Mei{\ss}ner}
\affiliation{Helmholtz-Institut~f\"{u}r~Strahlen-~und~Kernphysik~and~Bethe~Center~for~Theoretical~Physics,
~Rheinische~Friedrich-Wilhelms~Universit\"{a}t~Bonn,~D-53115~Bonn,~Germany}
\affiliation{Institute~for~Advanced~Simulation~(IAS-4),~Institut~f\"{u}r~Kernphysik~(IKP-3),
and~J\"{u}lich~Center~for~Hadron~Physics,~Forschungszentrum~J\"{u}lich,~D-52425~J\"{u}lich,~Germany}
\affiliation{Tbilisi State University, 0186 Tbilisi, Georgia}
\author{Bernard~Ch.~Metsch}%
\affiliation{Institute~for~Advanced~Simulation~(IAS-4),
~Forschungszentrum~J\"{u}lich,~D-52425~J\"{u}lich,~Germany}
\affiliation{Helmholtz-Institut~f\"{u}r~Strahlen-~und~Kernphysik,
~Rheinische~Friedrich-Wilhelms~Universit\"{a}t~Bonn,~D-53115~Bonn,~Germany}

\date{\today}%

\begin{abstract}
  We study the dependence of primordial nuclear abundances on
  fundamental nuclear observables such as binding energies, scattering
  lengths, neutron lifetime, \textit{etc.} by varying these
  quantities. The numerical computations were performed with four
  publicly available codes, thus facilitating an investigation of the
  model-dependent (systematic) uncertainties on these dependences.
  Indeed deviations of the order of a few percent are found. Moreover,
  accounting for the temperature dependence of the sensitivity of the
  rates to some relevant parameters leads to a reduction of the
  sensitivity of the final primordial abundances, which in some cases is
  appreciable. These effects are considered to be relevant for studies
  of the dependence of the nuclear abundances on fundamental parameters
  such as quark masses or couplings underlying the nuclear parameters
  studied here.
\end{abstract}

\maketitle

\section{\label{sec:intro}Introduction}
Primordial or Big Bang nucleosynthesis (BBN) is a fine laboratory to
test our understanding of nuclear and particle physics, for some
reviews see
Refs.~\cite{Olive:1999ij,Iocco:2008va,Cyburt:2015mya,Pitrou:2018cgg}.
The light elements generated in BBN can nowadays be calculated from
first principles, which offers the possibility to study in a
model-independent way the dependence of the element abundances as a
function of the fundamental parameters of the Standard Model of
particle physics, in particular the light quark masses $m_u, m_d$ and
the QCD $\theta$-parameter as well as the electromagnetic
fine-structure constant $\alpha$. This is based on the observation
that the strong and the electromagnetic interactions both contribute
to the nuclear binding, while the weak interactions makes their
presence feel via certain decays, most prominently in the neutron
decay into a proton and a lepton pair. In the Standard Model, the
quark masses are given in terms of \textit{a priori} unknown Yukawa
couplings to the Higgs boson and the strength (coupling constants) of
the various interactions must also be pinned down from experiment. It
is therefore obvious that the Standard Model should be an effective
field theory, which can eventually be derived from a more fundamental
theory, such as string theory or alike. Another possibility is the
Multiverse, in which our Universe is simply one particular
manifestation with the Yukawa couplings and interaction strengths
given as they are measured. Further, anthropic considerations come
into play by asking how much these parameters can be modified to still
allow for life on Earth (as given by the relative abundances of
certain elements like $^4$He, $^{12}$C and $^{16}$O without going into
details of galaxy and planet formation and alike).  For reviews on
these issues, see
\textit{e.g.} Refs.~\cite{Hogan:1999wh,Uzan:2002vq,Schellekens:2013bpa,Meissner:2014pma,Donoghue:2016tjk,Adams:2019kby}.

Here, we wish to revisit recent works that have derived bounds on the
fundamental parameters from element generation in BBN, see
\textit{e.g.} Refs.~\cite{Dent:2007zu,Berengut:2009js,Bedaque:2010hr,Berengut:2013nh,Lee:2020tmi}.
To really draw conclusions like that the Higgs vacuum expectation
value can only be varied by about 1\%
\cite{Bedaque:2010hr,Berengut:2013nh} when keeping the Yukawa
couplings constant, requires a full control of the systematic
uncertainties in the calculation of BBN nucleosynthesis within our
Universe. This is exactly what will be done in the present manuscript
which will serve as the basis for future studies of the allowed quark
mass and fine-structure constant variations within these
uncertainties. In particular, we study the dependence of the element
abundances on the nuclear reaction rates and also the temperature
dependence of the direct and radiative capture reactions in the BBN
network, that is most often neglected. Furthermore, variations in the
neutron lifetime, the singlet neutron-proton scattering length as well
as the deuteron binding energy are considered, where the latter two
enter the leading reaction $n+p\to d+\gamma$, which is particular
transparent using the effective field theory calculation of
Refs.~\cite{Chen:1999bg,Rupak:1999rk}.
Most importantly, we utilize four different
publicly available codes for BBN
\cite{Kawano:1992ua,Arbey:2011nf,Arbey:2018zfh,Pisanti:2007hk,Consiglio:2017pot,Gariazzo:2021iiu,Pitrou:2018cgg}
to address the systematic uncertainties related to the modeling of the
BBN network. In particular these codes differ in the number of nuclei
and reactions taken into account as well as in the specific
parametrization of the nuclear rates entering the coupled rate
equations for the BBN network. Moreover, in determining the
sensitivity of primordial abundances on nuclear parameters, we account
for a temperature dependence of the variation of some rates on
specific nuclear parameters and found that this leads in general to a
reduction of these sensitivities in comparison to previous studies,
where this temperature dependence was ignored.  To our knowledge, such
a comparative study has been not been published, but of course, new
cosmological results are always used to update BBN, see \textit{e.g.}
Ref.~\cite{Olive:2021noj}.

The manuscript is organized as follows: In Sect.~\ref{sec:bf} we
collect the basic formalism underlying BBN, which is required in what
follows.  In this section we also elaborate on the (pionless)
Effective Field Theory description of the leading $n+p \to d + \gamma$
reaction.  The BBN response matrix is introduced in
Sect.~\ref{sec:response}. The numerical results of this study are
presented in Sect.~\ref{sec:results} and discussed in
Sect.~\ref{sec:summary}.

\section{\label{sec:bf}Basic formalism}

The basic quantities to be determined in BBN are the nuclear
abundances $Y_{n_i}$, where $n_i$ denotes some nucleus.  As pointed
out \textit{e.g.} in
Refs.~\cite{Pitrou:2018cgg,Arbey:2018zfh,Gariazzo:2021iiu}\,,
the evolution of nuclear abundances $Y_{n_i}$ is given generically by
\begin{eqnarray}
  \dot Y_{n_1}
  &=&
      \sum_{\footnotesize
      \begin{array}[c]{c}
        n_2,\ldots,n_p\\
        m_1,\ldots,m_q\\
      \end{array}
  }
  N_{i_1}
  \Biggl(
  \Gamma_{m_1,\ldots,m_q \to n_1,\ldots n_p}
  \,\frac{Y_{m_1}^{N_{m_1}}\cdot Y_{m_q}^{N_{m_q}} }{
  N_{m_1}!\cdots N_{m_q}!}
  \nonumber\\ 
  &&
     \hspace*{5em}
     -\quad
     \Gamma_{n_1,\ldots,n_p \to m_1,\ldots m_q}
  \,\frac{Y_{n_1}^{N_{n_1}}\cdot Y_{m_q}^{N_{n_p}} }{
     N_{n_1}!\cdots N_{n_p}!}
     \Biggr)\,, \nonumber \\ &&
\end{eqnarray}
where the dot denotes the time derivative in a comoving frame,
$N_a$ is the stochiometric coefficient of species $a$ in the reaction
and where \textit{e.g.} for a two-particle reaction $a + b \to c + d$\,,
$\Gamma_{ab \to cd} = n_B \gamma_{ab\to cd}$ is the reaction rate with 
$n_B$ the baryon volume density\,.  This can readily be generalised to
reactions involving more (or less) particles,
see~\cite{Pitrou:2018cgg}\,. These equations are coupled via corresponding
energy densities to  the standard Friedmann equation describing the cosmological
expansion in the early universe, for details and basic assumptions,  see
also~\cite{Pitrou:2018cgg,Arbey:2018zfh,Gariazzo:2021iiu}.

\subsection{\label{subsec:nrr}Nuclear reaction rates}

The average reaction rate $\gamma_{ab\to cd} = N_A\,\langle
\sigma_{ab\to cd}\,v\rangle$ for a two-particle reaction $a + b \to c
+ d$ is obtained by folding the cross section $\sigma_{ab\to cd}(E)$
with the Maxwell-Boltzmann velocity distribution in thermal
equilibrium
\begin{equation}
  \label{eq:crossrate}
  \gamma_{ab\to cd}(T)
  =
  N_A\,
  \sqrt{\frac{8}{\pi \mu_{ab} (kT)^3}}\,
  \intdif{0}{\infty}{E}\,E\,\sigma_{ab\to cd}(E)\,\expo{-\frac{E}{kT}}\,,
\end{equation}
conventionally multiplied by Avogadro's number $N_A$\,, where
$\mu_{ij}$ is the reduced mass of the nuclide pair $ij$\,, $E$ is the
kinetic energy in the center-of-mass (CMS) system, $T$ is the
temperature and $k$ the Boltzmann constant.  Defining $y := {E}/(k T)$
this can be written in the form
\begin{eqnarray}
  \gamma _{ab\to cd}(T)
  &=& N_A
      \sqrt{\frac{8\,k T}{\pi\,\mu_{ab}}}
      \intdif{0}{\infty}{y}\,\sigma_{ab\to cd}(k T\,y)\,y\,\expo{-y},
\end{eqnarray}
which is suited for numerical computation \textit{e.g.} with a
Gauss-Laguerre integrator.

With the detailed balance relation
\begin{equation}
  \sigma_{cd\to ab}(E') 
  =
  \frac{g_a\,g_b}{g_c\,g_d}\,\frac{p^2}{{p'}^2}\,\sigma_{ab\to cd}(E)\,,
\end{equation}
where 
\begin{equation}
E = \frac{p^2}{2\,\mu_{ab}}\,, ~~~E'=\frac{{p'}^2}{2\,\mu_{cd}}~,
\end{equation}
are the kinetic energies in the entrance and exit channels,
respectively, and $g_i$ is the spin multiplicity of particle $i$.
Energy conservation leads to,
\begin{eqnarray}
  m_a + m_b + E
  &=&
      m_c + m_d + E'
      \quad
      \textnormal{or}
      \quad
      \nonumber  \\
  E' = E + Q\,,&\quad\textnormal{with}\quad&
                                             Q = m_a+m_b-m_c-m_d~,
\end{eqnarray}
in terms of the $Q$-value for the forward reaction.  The inverse
reaction  rate is related to the forward rate as
\begin{equation}
  \label{eq:inverserate}
  \gamma_{cd \to ab}(T)
  =
  \left(\frac{\mu_{ab}}{\mu_{cd}}\right)^{\frac{3}{2}}
  \frac{g_a\,g_b}{g_c\,g_d}\, 
  \expo{-\frac{Q}{k T}}\,
  \gamma_{ab\to cd}(T)\,.
\end{equation}

\subsection{$Q$-value dependence of the reaction rates}

The rates for nuclear reactions of the kind
\begin{equation}
  a + b \to c + d
\end{equation}
depend on the $Q$-value of the reaction, which in turn depends on the
nuclear binding energies.

\subsubsection{Varying nuclear binding energies}

Now, \textit{e.g.} for a relative change in the binding energy of
nucleus $a: B_a \mapsto B_a\,(1+\delta_a)$ the nuclear mass change is
given by
\begin{equation*}
  \Delta m_a = m_a - \delta_a\,B_a
\end{equation*}
and thus the reduced mass occurring in the expressions above changes as 
\begin{eqnarray}
  \mu_{ab}   
  &\mapsto&
            \frac{(m_a-\delta_a\,B_a)\,m_b}{m_a - \delta_a\,B_a + m_b}
            =
            \frac{m_a\,m_b-\delta_a\,B_a\,m_b}{(m_a+m_b)
            \left(1-\frac{\delta_a\,B_a}{m_a+m_b}\right)}
            \nonumber  \\
  &\approx&
            \left(\mu_{ab} - \frac{\delta_a\,B_a\,m_b}{m_a+m_b}\right)
            \left(1+\frac{\delta_a\,B_a}{m_a+m_b}\right)
            \nonumber  \\
  &\approx&
            \mu_{ab}\left(1+\frac{\delta_a\,B_a}{m_a+m_b}
            -\frac{\delta_a\,B_a}{m_a}\right)
 \nonumber \\
  &=&
      \mu_{ab}\left(1-\frac{\delta_a\,B_a\,m_b}{(m_a+m_b)\,m_a}\right)
      =
      \mu_{ab}\left(1-\mu_{ab}\,\frac{\delta_a\,B_a}{m_a^2}\right)\,.
      \nonumber\\ &&
\end{eqnarray}
Since we shall consider fractional changes in the binding energy
$\delta_a \approx \mathcal{O}(10^{-3})$ and ${B_a}/{m_a} \approx
\mathcal{O}(10^{-2}-10^{-3})$\,, the change in the reduced masses is
very small, $\mathcal{O}(10^{-5}-10^{-6})$ and this change in the
reduced masses will therefore be neglected.

Noting that, since the total number of protons and neutrons is
conserved in this kind of reaction: $Z_a+Z_b = Z_c + Z_d\,, N_a + N_b
= N_c + N_d$,
\begin{eqnarray}
  Q
  &=&
      m_a + m_b - m_c - m_d
 \nonumber \\
  &=&
      Z_a\,m_p + N_a\,m_n - B_a
      +
      Z_b\,m_p + N_b\,m_n - B_b
 \nonumber \\
  &&\quad
     -
     Z_c\,m_p - N_c\,m_n + B_c
     -
     Z_d\,m_p - N_d\,m_n + B_d
 \nonumber \\
  &=&
     B_c + B_d - B_a - B_b\,, 
\end{eqnarray}
a change in the binding energy will only affect the $Q$ value:
\begin{equation}
  \Delta Q = -\delta_a\,B_a - \delta_b\,B_b + \delta_c\,B_c + \delta_d\,B_d\,,
\end{equation}
where $\delta_i$ is the fractional change in the binding energy $B_i$
of nuclide $i$ and this will be the major effect to be studied below.
We also note that in case of three-particle final states we shall use
the same formula, \textit{e.g.} by taking
${}^{4}\textnormal{He}+{}^{4}\textnormal{He}$ as
${}^{8}\textnormal{Be}$ and $p+n$ as $d$ \textit{etc}.\ .

Note that, apart from the reaction $n + p \to d + \gamma$, to be
discussed in some detail in Sect.~\ref{sss:npdg}, we do not make any
assumptions concerning the binding energy dependence of the cross
section itself, although, as has been argued in~\cite{Dent:2007zu}, 
\textit{e.g.}  a $B_d^{-1}$ dependence for reactions involving
deuterons might be considered.

\subsubsection{\label{sss:dr}Direct reactions $a + b \to c + d$:}

Since the cross section is proportional to the final channel momentum
$p'$ and the probability of two charged nuclear particles overcoming
their electrostatic barriers is given by the so called
Gamow-Sommerfeld-factor~\cite{gamow1928}
\begin{equation}
  \expo{-\sqrt{\frac{E'_G}{E'}}}\,,
  \quad\textnormal{where}\quad
  E'_G = 2\pi^2\,Z_c^2\,Z_d^2\,\alpha^2\,\mu_{cd}\,c^2
\end{equation}
is the so called Gamow-energy for the exit channel, $Z_i$ is the
charge number of nuclide $i$ and $\alpha \simeq 1/137$ is the
electromagnetic fine-structure constant, the $Q$-value dependence is
given by
\begin{equation}
  \sigma(Q;E)_{a b \to c d}
  \propto
  \sqrt{Q+E}\,\expo{-\sqrt{\frac{E'_G}{Q+E}}}\,.
\end{equation}
One thus finds for a change in the $Q$-value
\begin{equation}
  \widetilde Q = Q_0 + \Delta Q
\end{equation}
with 
\begin{equation}
  \left(\diff{\sigma}{Q}\right)(Q_0;E)
  =
  \frac{1}{2}\,\sigma(Q_0;E)\left(
    1 + \sqrt{\frac{E'_G}{Q+E}}\right)\,\frac{1}{Q+E}
\end{equation}
that
\begin{eqnarray}
  \lefteqn{
  \sigma_{ab\to cd}(\widetilde Q;E)
  \approx
  \sigma_{ab\to cd}(Q_0;E)
  +
  \left(\diff{\sigma}{Q}(Q_0;E)\right)\,\Delta Q
  }
  \nonumber\\
&=&
    \sigma_{ab\to cd}(Q_o;E)
    \left(
    1
    +
    \frac{\Delta Q}{2\,(Q_0+E)}
    \left(1+\sqrt{\frac{E'_G}{Q_0+E}}\right)
    \right)\,. \nonumber\\ &&
\end{eqnarray}
Thus, in linear approximation we have
\begin{eqnarray}
  \label{eq:Qdepx}
  \gamma_{ab\to cd}(\widetilde Q;T)
  &\approx&
            \gamma_{ab\to cd}(Q_0; T)
            \nonumber\\
  &&\hspace*{-5em}
     +\,\,
     \sqrt{\frac{8}{\pi\,\mu_{ab}\,(kT)^3}}\,
     \intdif{0}{\infty}{E}\,E\,\sigma(Q_0;E)
     \nonumber\\
  &&
     \left(
     \frac{\Delta Q}{2\,(Q_0+E)}
     +
     \frac{\Delta Q\,\sqrt{E'_G}}{2\,(Q_0+E)^\frac{3}{2}}
     \right)\,\expo{-\frac{E}{kT}}
     \nonumber\\
  &=:&
       \gamma_{ab\to cd}(Q_0; T) + \Delta\gamma(T)_{ab\to cd}
\end{eqnarray}  
and the determination of the temperature dependent $Q$-value change
requires a separate computation of an integral over the cross section
or, equivalently, over the astrophysical $S$-factor defined as
\begin{equation}
  S(E) := \sigma(E)\,E\,\expo{\sqrt{\frac{E_G}{E}}}\,.
\end{equation}
We shall use a numerical integration of Eq.~(\ref{eq:Qdepx}) to
determine the $Q$-dependence of the rates for a dozen leading
reactions in the BBN network.

If, however, we assume that for the relevant energies in the integrand
$E \ll Q_0$, which might be the case for low temperatures, then this
simplifies to
\begin{eqnarray}
  \label{eq:Qdepa}
  \gamma_{ab\to cd}(\widetilde Q; T)
  &\approx&
            \gamma_{ab\to cd}(Q_0; T)
            \nonumber\\
  &&\hspace*{1.25em}
     \left(1 + \frac{\Delta Q}{2\,Q_0}
     \left(1 + \sqrt{\frac{E'_G}{Q_0}}\right)
     \right)
     \nonumber\\
  &&\hspace*{-8em}=
     \gamma_{ab\to cd}(Q_0; T)
     \left(1 + \frac{1}{2}\,\delta_Q\,
     \left(1 + \sqrt{\frac{E'_G}{Q_0}}\right)
     \right)
     \nonumber\\
  &=:&
       \gamma_{ab\to cd}(Q_0; T) + \Delta\gamma_{0;ab\to cd}\,,
\end{eqnarray}
where $\delta_Q = {\Delta Q}/{Q_0}$ is the fractional change in
the $Q$-value.  Indeed, Eq.~(\ref{eq:Qdepa}) is the approximation that
has been used in previous studies, such as \textit{e.g.}
Refs.~\cite{Dent:2007zu,Berengut:2009js,Bedaque:2010hr,Berengut:2013nh}.
Note that with this approximation the $Q$-value change is temperature
independent.

\subsubsection{\label{sss:rc}Radiative capture reactions: $a + b \to c + \gamma$:}

Assuming that the radiation is dominated by electromagnetic dipole transitions%
\footnote{%
  Note, however, that this not always the case, exceptions with
  appreciable $E2$ (electric quadrupole) contributions are
  \textit{e.g.} the reactions:
  ${^2}\textnormal{H}+{^2}\textnormal{H} \to {^4}\textnormal{He} + \gamma$\,,
  ${^2}\textnormal{H}+{^4}\textnormal{He} \to {^6}\textnormal{Li} + \gamma$ and
  ${^4}\textnormal{He}+{^{12}}\textnormal{O} \to {^{16}}\textnormal{O} + \gamma$\,.
  We nevertheless always assume dipole dominance.
}%
, we have
\begin{equation}
  \sigma(Q;E)_{ab\to c\gamma}
  \propto
  E_\gamma^3 \approx (Q+E)^3\,.
\end{equation}
Accordingly, for a change in the $Q$-value
\begin{equation}  
  \widetilde Q = Q_0 + \Delta Q
\end{equation}
with
\begin{equation}
  \left(\diff{\sigma}{Q}\right)(Q_0;E)
  =
  3\,\frac{\sigma(Q_0;E)}{Q_0+E}
\end{equation}
one finds that
\begin{eqnarray}
  \sigma_{ab\to c\gamma}(\widetilde Q;E)
  &\approx&
            \sigma_{ab\to c\gamma}(Q_0;E)
            +
            \left(\diff{\sigma}{Q}(Q_0;E)\right)\,\Delta Q
            \nonumber  \\
  &=&
      \sigma_{ab\to c\gamma}(Q_o;E)
      \left(
      1+3\,\frac{\Delta Q}{Q_0+E}
      \right)
\end{eqnarray}
and thus the change in the rate
\begin{eqnarray}
  \label{eq:radcapx}
  \gamma_{ab\to c\gamma}(\widetilde Q; T)
  &\approx&
            \gamma_{ab\to c\gamma}(Q_0; T)
            \nonumber\\
  &&\hspace*{-7em}
     +\,
     3\,\Delta Q\,
     \sqrt{\frac{2}{\pi\,\mu_{ab}\,c^2\,(kT)^3}}\,
     \intdif{0}{\infty}{E}\,\,\sigma(Q_0;E)
     \nonumber\\
  &&\hspace{6em}
     \frac{E}{Q_0+E}
     \expo{-\frac{E}{kT}}
     \nonumber\\
  &=:&
       \gamma_{ab\to c\gamma}(Q_0; T) + \Delta\gamma(T)_{ab\to c\gamma}\,,
\end{eqnarray}
as for the direct reactions before is temperature dependent. 

If again, as in
Refs.~\cite{Dent:2007zu,Berengut:2009js,Bedaque:2010hr,Berengut:2013nh},
we assume that for the relevant energies in the integrand $E \ll Q_0$
then this simplifies to
\begin{eqnarray}
  \label{eq:radcapa}
  \gamma_{ab\to c\gamma}(\widetilde Q; T)
  &=&
      \gamma_{ab\to c\gamma}(Q_0; T)
      \left(1 + \frac{3\,\Delta Q}{Q_0}\right)
      \nonumber\\
  &=&
      \gamma_{ab\to cd}(Q_0; T)
      \left(1 + 3\,\delta_Q\right)
      \nonumber\\
  &=:&
       \gamma_{ab\to c\gamma}(Q_0; T) + \Delta\gamma_{0;ab\to c\gamma}\,,
\end{eqnarray}
where $\delta_Q := {\Delta_Q}/{Q_0}$\,. 
In this case:
\begin{equation}
  Q = m_a + m_b - m_c = B_c - B_a - B_b
\end{equation}
and a fractional change in the binding energy: $B_i \mapsto ( 1 +
\delta_i\,B_i)$ will be assumed, as above, to affect the $Q$ value
only:
\begin{equation}
  \Delta Q = -\delta_a\,B_a - \delta_b\,B_b + \delta_c\,B_c\,.
\end{equation}

\subsubsection{\label{sss:wd}Weak decay rates $a \to b + e^{\pm} + \stackrel{(-)}{\nu}$}

Neglecting the nuclear recoil of the daughter nucleus, we have from
energy conservation
\begin{equation}
  M_a = M_b + E_e + E_\nu
\end{equation}
and the maximum $e^\pm$ energy is given by
\begin{equation}
  E_e^{\footnotesize{max}} = M_a - M_b =: Q\,
\end{equation}
where $M_a\,,M_b$ are the nuclear parent and daughter masses.
Accordingly, the maximal $e^\pm$ momentum is given by
\begin{equation}
  p_e^{\footnotesize{max}} = \sqrt{Q^2 - m_e^2} = m_e\,\sqrt{q^2-1}\,,
  \quad q := \frac{Q}{m_e}\,.
\end{equation}
The weak decay rate then reads~\cite{es1977}\,:
\begin{equation}
  \lambda
  =
  \frac{G^2}{2\pi^3}\,\frac{m_p\,c^2}{\hbar}|M_{if}|^2\,f(Z,q)\,,
\end{equation}
where $G$ is the weak (Fermi) coupling constant, $M_{if}$ the nuclear
matrix element and (ignoring Coulomb corrections):
\begin{eqnarray}
  f(q) 
  &:=&
       \left.f(Z,q)\right|_{Z=0}
       \nonumber\\
  &=&
      \frac{1}{60}\Bigl[
      \sqrt{q^2-1}\,(2\,q^4-9\,q^2-8)
      \nonumber \\
  &&\hspace*{5em}
     + 15\,q\,\log(q+\sqrt{q^2-1})
     \Bigr]\,.
\end{eqnarray}
From this an expression for the fractional change in the rate due to a
change in the $Q$-value,
\begin{eqnarray}
  \label{eq:wdlogf}
  \diff{\log(f(q))}{q}
    &=&
        \nonumber\\
    &&\hspace*{-4em}
       \frac{
       \sqrt{q^2-1}\,(10\,q^3-25\,q) + 15\,\log(q+\sqrt{q^2-1})
       }{
       \sqrt{q^2-1}\,(2\,q^4-9\,q^2-8) + 15\,q\,\log(q+\sqrt{q^2-1})
       } \nonumber\\ &&
\end{eqnarray}
follows, \textit{i.e.}
\begin{equation}
  \label{eq:wdf}
  \diff{\log(\lambda)}{q}
  =
  \frac{1}{\lambda}\,\diff{\lambda}{q}
  =
  \diff{\log{(f(q))}}{q}\,.
\end{equation}
With $\widetilde Q = Q_0 + \Delta Q = Q_0\,(1 + \delta_Q)$\,,
$\lambda_0 := \lambda(Q_0)$ and $q_0 := {Q_0}/{m_e}$, we find
\begin{eqnarray}
  \label{eq:wdx}
  \widetilde{\lambda}
  &=&
      \lambda(Q_0) + \Delta \lambda
      \approx
      \lambda(Q_0) + \diff{\lambda}{Q}(Q_0)\,\Delta Q
      \nonumber\\
  &=&
      \lambda(Q_0) + \diff{\lambda}{q}\,q_0\,\delta_Q
      =
      \lambda(Q_0) + \lambda_0\,\diff{\log(\lambda)}{q}\,q_0\,\delta_Q
      \nonumber\\
  &=&
      \lambda(Q_0)\left(1+q_0\,\diff{\log(f(q))}{q}\,\delta_Q\right)~.
\end{eqnarray}

\subsubsection{\label{sss:npdg}The leading reaction $n + p \to d + \gamma$}
 
The formulas given above are the basis for the calculation of the
primordial abundance variations mainly due to the dependence on the
$Q$-values of most reactions in the BBN-network.

For the leading reaction $n + p \to d + \gamma$ in this network it is
possible to study more details, since for this reaction an accurate
description for the relevant energies in the framework of the
so-called pionless effective field theory~\cite{Chen:1999bg,Rupak:1999rk}
is available. According to that work, the cross section for the capture
reaction $n + p \to d + \gamma$ is given by
\begin{equation}
  \sigma_{np\to d\gamma}(E)
  =
  \frac{4\pi\,\alpha\,(\gamma^2+p^2)^3}{\gamma^3\,M_N^4\,p}
  \left[
    \left|\widetilde{\chi}_{M1}\right|^2
    +
    \left|\widetilde{\chi}_{E1}\right|^2
    \right]\,
\end{equation}
with $\alpha = {e^2}/{\hbar c}$ the electromagnetic fine-structure
constant and $E={p^2}/{M_N}$ is the total kinetic energy in the CMS,
with $p$ the magnitude of the momentum of either the proton or the
neutron, and $\gamma := \sqrt{B\,M_N}$ is the ``binding momentum'' of
the deuteron ground state.\footnote{%
  Here it is understood that all energies are expressed through
  equivalent wave numbers, \textit{i.e.} $p \mapsto {p\,c}/({\hbar
    c})$\,, $M \mapsto {M\,c^2}/({\hbar c})$\,, $\gamma \mapsto
  {\sqrt{B\,M\,c^2}}/({\hbar c})$\,, \textit{etc.}\,, all in units of
  $\textnormal{fm}^{-1}$\,. The cross section is then given in units of
  $\textnormal{fm}^2 = 10\,\textnormal{mb}$\,.
}
The (dimensionless) electric dipole contribution is given up to
N${}^3$LO by
\begin{eqnarray}
  \left|\widetilde{\chi}_{E1}\right|^2
  &=&
      \frac{\gamma^4\,M_N^2\,p^2}{(\gamma^2+p^2)^4}
      \Biggl[
      \underbrace{
      1+\gamma\,\rho_d+(\gamma\,\rho_d)^2+(\gamma\,\rho_d)^3}
      _{\approx \frac{1}{1-\gamma\,\rho_d}}
      \nonumber  \\
  &+&
      \frac{\gamma\,M_N}{6\pi}\left(\frac{\gamma^2}{3}+p^2\right)
      \nonumber  \\
  &\times&
           \left(
           {}^{\slash{\!\!\!\pi}}C_2^{({}^3 P_0)}
           +
           2\,{}^{\slash{\!\!\!\pi}}C_2^{({}^3 P_1)}
     +
           \frac{20}{3}{}^{\slash{\!\!\!\pi}}C_2^{({}^3 P_2)}
           \right)
           \Biggr],
\end{eqnarray}
where $\rho_d$ is the effective range. We use the numerical values 
from Ref.~\cite{Chen:1999bg} ,
$\gamma^{-1}=4.318\,946\,\,\textnormal{fm}$,
$\rho_d=1.764\pm0.002\,\,\textnormal{fm}$,
${}^{\slash{\!\!\!\pi}}C_2^{({}^3 P_0)}=6.53\,\,\textnormal{fm}^4$,
${}^{\slash{\!\!\!\pi}}C_2^{({}^3 P_1)}=-5.91\,\,\textnormal{fm}^4$,
${}^{\slash{\!\!\!\pi}}C_2^{({}^3 P_0)}=0.57\,\,\textnormal{fm}^4$\,.
Furthermore, the magnetic dipole contribution at NLO reads
\begin{eqnarray}
  \left|\widetilde{\chi}_{M1}\right|^2
  &=&
      \frac{\gamma^4\,\kappa_1^2\,(1-a_s\,\gamma)^2}{
      (1+a_s^2\,\gamma^2)(\gamma^2+p^2)^2}
 \nonumber \\
  &\times &
            \Biggl[
            1 
            +
            \gamma\,\rho_d
            -
            \frac{r_0}{a_s}\,\frac{(\gamma\,a_s+a_s^2\,p^2)\,a_s^2\,p^2}{(1+a_s^2\,p^2)(1-a_s\,\gamma)}
            \nonumber  \\
  &&
     \hspace*{5em}
     -
     \frac{{}^{\slash{\!\!\!\pi}}L_{np}}{\kappa_1}\,\frac{a_s\,M_N}{2\pi}\,\frac{\gamma^2+p^2}{1-a_s\,\gamma}
     \Biggr]\,,
\end{eqnarray}
where $\kappa_1 = (\mu_p-\mu_n)/2$ is the nucleon isovector magnetic
moment (in units of $\mu_N = {e\,\hbar}/(M_N\,c)$)\,, $a_s$ is the
scattering length and $r_0$ is the effective range in the
${}^1S_0$-channel.  Again we use the numerical values from
Ref.~\cite{Chen:1999bg} ,
  $a_s = -23.714\pm 0.013\,\,\textnormal{fm}$,
  $r_0 = 2.73\pm0.03\,\,\textnormal{fm}$,
  ${}^{\slash{\!\!\!\pi}}L_{np} = -4.513\,\,\textnormal{fm}^2$\,.
The resulting cross section is compared to experimental data in
Fig.~\ref{{fig:npdg}}\,.

\begin{figure}[htb] 
  \includegraphics[width=\columnwidth,viewport = 320 470 620 720]{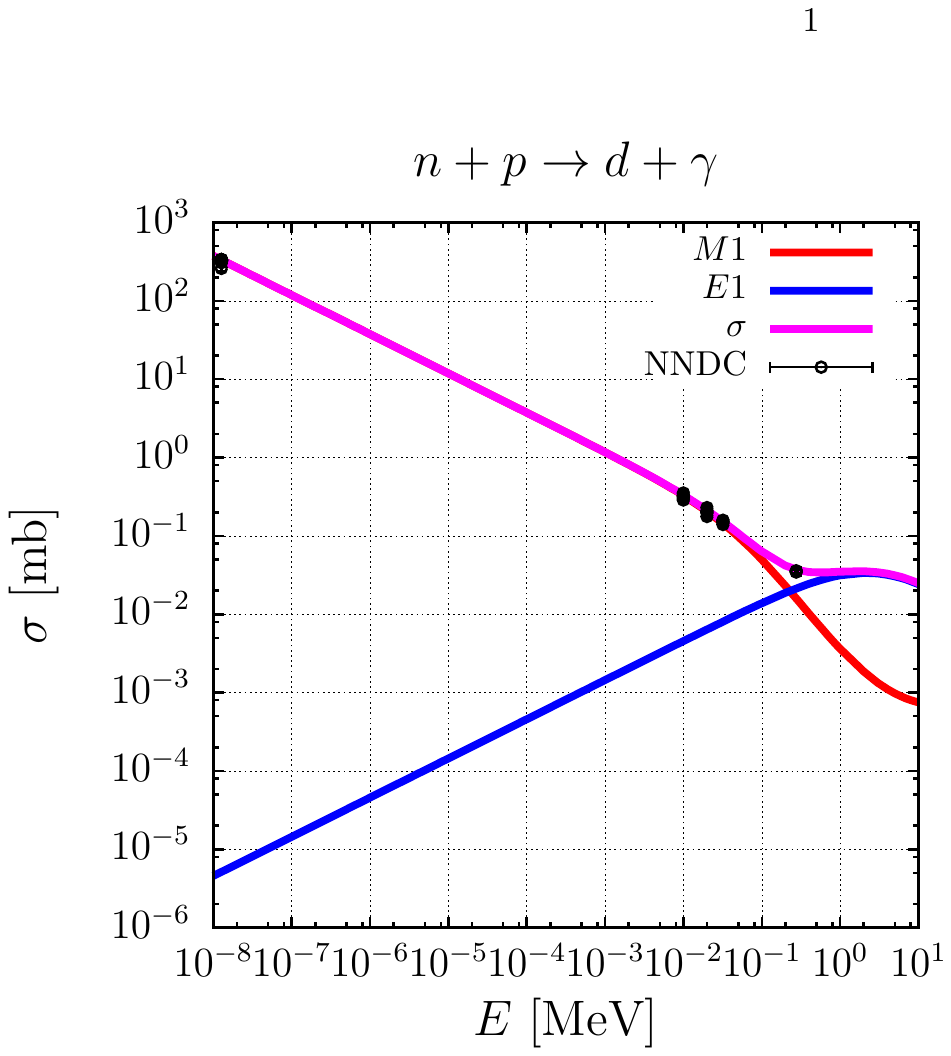}
  \caption{\label{{fig:npdg}}%
    Radiative capture $n + p \to d + \gamma$ cross section (in mb)
    (magenta, color online) as a function of the CMS kinetic energy in
    MeV, from~\cite{Rupak:1999rk}\,; the calculated $M1$ (red) and $E1$
    (blue) contributions to the cross section are also shown
    separately. Experimental data later than 1960 are from NNDC
    online~\cite{nndc}\,.
  }
\end{figure} 

Considering only the leading contributions one finds
\begin{equation}
  \sigma_{M1;np\to d\gamma}(p)
  =
  \frac{
    \pi\,\alpha\,(\mu_n-\mu_p)^2\,\gamma\,(\gamma^2+p^2)\,(1-a_s\,\gamma)^2
  }{
    M_N^4\,p\,(1+a_s^2\,p^2)
  }\,.
\end{equation}
With $p=\sqrt{M_N\,E}, \gamma=\sqrt{M_N\,B_d}$ ($B_d$ the binding
energy of the deuteron) and introducing $W$ via $a_s=:
-1/\sqrt{M_N\,W}$ this reads
\begin{eqnarray}
  \sigma_{M1;np\to d\gamma}(E)
  &=&
  \pi\,\frac{\alpha}{M_N^2}\,(\mu_n-\mu_p)^2
  \nonumber\\
  &&\hspace*{-2em}\times
  \frac{
  \sqrt{B_d}\,(B_d + E)\left(\sqrt{B_d}+\sqrt{W}\right)^2
  }{
  M_N\,\sqrt{E}\left(E+W\right)
  }.
\end{eqnarray}
Likewise
\begin{equation}
  \sigma_{E1;np\to d\gamma}(p)
  =
  \frac{4\pi\,\alpha\,p\,\gamma}{
    M_N^2\,(\gamma^2+p^2)\,(1-\gamma\,\rho_d)
  }
\end{equation}
or, with $\rho_d =: {1}/(\sqrt{M_N\,R})$\,,
\begin{equation}
  \sigma_{E1;np\to d\gamma}(E)
  =
  4\pi\,\frac{\alpha}{M_N^2}\,
  \frac{
    \sqrt{E\,B_d}
  }{
    (E + B_d)\left(1-\sqrt{\frac{B_d}{R}}\right)
  }\,.
\end{equation}

The cross section for the inverse reaction can be obtained via
\begin{eqnarray}
  \sigma_{\gamma\,d\to np}(E_\gamma)
  &=&
      \frac{2}{3}\,\frac{p^2}{k^2}\,\sigma_{np\to d\gamma}(E_\gamma-B_d)
\nonumber  \\
  &&\hspace*{-5em}=
      \frac{2}{3}\,
      \frac{M_N c^2\,(E_\gamma-B)_d}{E_\gamma^2}\,
      \sigma_{np\to d\gamma}(E_\gamma-B_d)
\end{eqnarray}
with $E_\gamma = K$ the photon energy in the rest frame of the deuteron,
where we used $k\approx K$ and 
\begin{eqnarray}
  p^2
  &\approx&
            M_N\,(k-B_d) \approx M_N\,(K-B_d)
            =
            M_N\,k - \gamma^2
\nonumber  \\
  &&
  \,\,\Leftrightarrow\,\,
  p^2+\gamma^2 \approx M_N\,k \approx M_N\,K\,
\end{eqnarray}
as well as the ratio of the statistical weights 
\begin{equation}
  \frac{g_p\,g_n}{g_\gamma\,g_d}
  =
  \frac{2\cdot 2}{2\cdot 3} = \frac{2}{3}\,.
\end{equation}
and
\begin{equation}
  E
  =
  \frac{p^2}{M_N} \approx k - B_d \approx K - B_d = E_\gamma - B_d\,.
\end{equation}

\begin{figure}[htb] 
  \includegraphics[width=\columnwidth,viewport = 320 470 620 720]{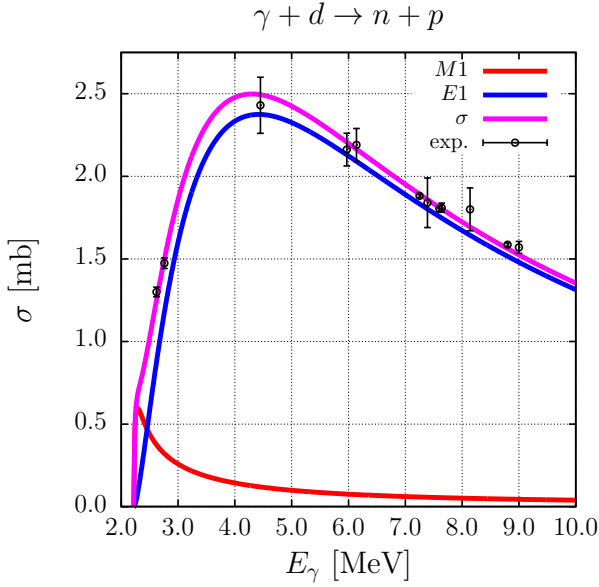}
  \caption{\label{fig:gdnp}%
    Photodisintegration $\gamma + d \to n + p$ cross section (in mb)
    (magenta, color online) as a function of the (laboratory) photon
    energy in MeV, from~\cite{Rupak:1999rk}.  The calculated $M1$ (red) and
    $E 1$ (blue) contributions to the cross section are also shown
    separately. Experimental data from Ref.~\cite{as1991}. 
  }
\end{figure}

We thus find for the photodisintegration reaction
\begin{eqnarray}
  \sigma_{M1;\gamma d\to np}(p)
  &&
\nonumber  \\
  &&\hspace*{-3em}=
      \frac{2\pi}{3}\,\alpha\,\frac{(\mu_n-\mu_p)^2\,p^2\,\gamma\,(\gamma^2+p^2)^2\,(1-a_s\,\gamma)^2}
      {k^2\,M_N^4\,p\,(\gamma^2+p^2)\,(1+a_s^2\,p^2)}
\nonumber  \\
  &&\hspace*{-3em}=
      \frac{2\pi}{3}\,\frac{\alpha}{M_N^2}\,(\mu_n-\mu_p)^2\,
      \frac{p\,\gamma\,(1-a_s\,\gamma)^2}{
      (\gamma^2+p^2)\,(1+a_s^2\,p^2)}
\nonumber  \\
  &&\hspace*{-6em}=
      \frac{2\pi}{3}\,\frac{\alpha}{M_N^2}\,(\mu_n-\mu_p)^2\,
      \frac{\sqrt{B_d}\,\sqrt{E_\gamma-B_d}\,(\sqrt{B_d}+\sqrt{W})^2}{E_\gamma\,(E_\gamma-B_d+W)}
\nonumber  \\
  &=&
      \sigma_{M1;\gamma d\to np}(E_\gamma)\,,
\end{eqnarray}
where we used $W = (M_N\,a_s^2)$\,.
This indeed corresponds to the Bethe-Longmire zero range
expression, see Eq.~(56) of Ref.~\cite{Bethe:1950jm}. For the electric
dipole contribution to the photodisintegration cross section one finds
the zero-range expression
\begin{eqnarray}
  \sigma_{E1;\gamma d\to np}(p)
  &&
   \nonumber   \\
  &&\hspace*{-3em}=
      \frac{8\pi}{3}\,\alpha\,
      \frac{p^3\,\gamma\,(\gamma^2+p^2)^2}{k^2\,M_N^2\,(\gamma^2+p^2)^3
      }
  \nonumber \\
  &&\hspace*{-3em}=
      \frac{8\pi}{3}\,\alpha\,
      \frac{\gamma\,p^3}{(\gamma^2+p^2)^3}
      =
      \frac{8\pi}{3}\,\frac{\alpha}{\gamma^2}\,\frac{(p\,\gamma)^3}{(\gamma^2+p^2)^3}
  \nonumber \\
  &&\hspace*{-3em}=
      \frac{8\pi}{3}\,\frac{\alpha}{\gamma^2}\,
      \frac{M_N^3\,B_d^3\,(1-\eta)^{\frac{3}{2}}}{M_N^3\,E_\gamma^3}
      =
      \frac{8\pi}{3}\,\frac{\alpha}{\gamma^2}\,\frac{(1-\eta)^{\frac{3}{2}}}{\eta^3}
  \nonumber \\
  &&\hspace*{-3em}=
      \frac{8\pi}{3}\,\frac{\alpha}{M_N}\,\frac{\sqrt{B_d}\,(E_\gamma-B_d)^{\frac{3}{2}}}{E_\gamma^3}
  \nonumber\\
     &&\hspace*{-3em}=
      \sigma_{E1;\gamma d\to np}(E_\gamma)
\end{eqnarray}
where $\gamma^2 + p^2\approx M_N\,k \approx M_N\,K =: M_N\,E_\gamma$
and  $\eta := {E_\gamma}/{B_d}$\,, corresponding to the
expression of Eq.~(16) in~\cite{bp1935}, first derived by Bethe and
Peierls\,. The effective range correction (see Eq. (6)
in~\cite{Bethe:1950jm}) to this is given by the factor
$(1-\gamma\,\rho_d)^{-1}$\,. The result from the N${}^3$LO-calculation
of~\cite{Rupak:1999rk} is compared to experimental data in
Fig.~\ref{fig:gdnp}\,.

\begin{figure}[htb] 
  \includegraphics[width=\columnwidth,viewport = 320 470 620 720]{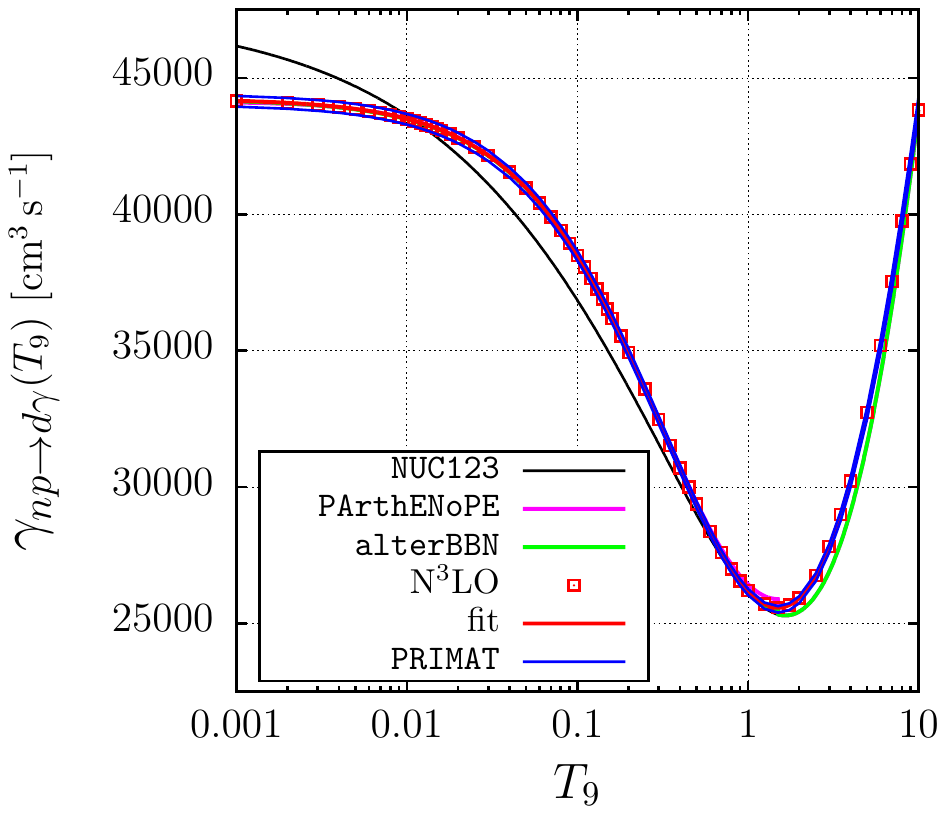}
  \caption{\label{fig:rate_npdg}%
    The temperature dependent rate (in units of cm${}^3$\,s${}^{-1}$)
    calculated from the cross section of the $n + p \to d + \gamma$
    reaction in N${}^3$LO and compared to the parameterizations used in
    the programmes \texttt{NUC123}~\cite{Kawano:1992ua} (in black, color online),
    \texttt{PArthENoPE}~\cite{Gariazzo:2021iiu} (in magenta),
    \texttt{AlterBBN}~\cite{Arbey:2018zfh} (in green) 
    and \texttt{PRIMAT}~\cite{Pitrou:2018cgg} for which the two blue curves
    correspond to the lower and upper limit of the rate.
    The red solid curve represents a rational function fit to the red data points.
    $T_9 := T / [10^9\,\textnormal{K}]$\,. 
}
\end{figure}

Therefore, for this leading reaction in the BBN-network a sufficiently
accurate description of the experimental data is available. Moreover, this
EFT-description allows to study the variation of the cross section due
to changes in nuclear parameters, such as the deuteron binding energy
$B_d$, the ${}^1 S_0$ $np$ scattering length $a_s$, the corresponding
effective range $r_0$ as well as the effective range $\rho_d$ in the
deuteron channel of  $np$-scattering. Below we shall concentrate on
the dependence on $B_d$ and $a_s$, since these give the leading effects.

In Fig.~\ref{fig:rate_npdg} the numerically calculated rate according
to Eq.~(\ref{eq:crossrate}) for the $n + p \to d + \gamma$ reaction is
shown and compared to the parameterizations used in the original
version of the codes. Apart from the original \texttt{NUC123}
parametrisation, see~\cite{Smith:1992yy}, these curves are very
consistent. In the rest of this paper, the parameterization according to a
rational function fit of the form
\begin{equation}
  f(t) = c_0¸\,\frac{1+\sum_{i=1}^5 c_i\,t^i}{1+\sum_{i=1}^4 d_j\,t^j}
\end{equation}
to the calculated rate based on the N${}^3$LO-cross section is used
throughout in all programmes. 

\begin{figure}[hbt] 
  \includegraphics[width=\columnwidth,viewport = 320 470 620 720]{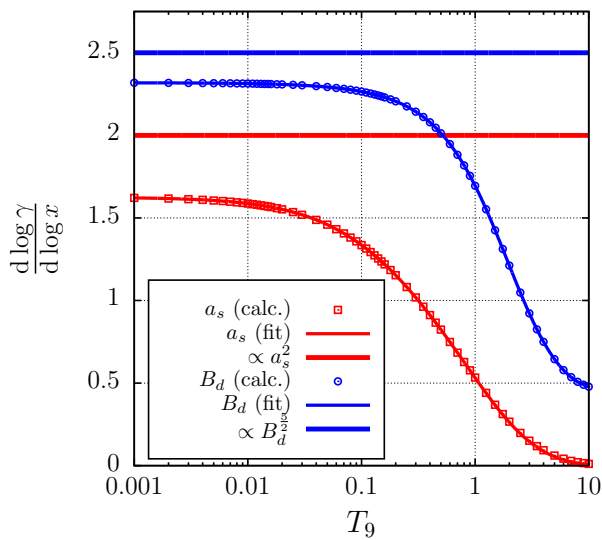}
  \caption{\label{fig:dloggdlogx}%
    Temperature dependence of the fractional change in the rate
    $\gamma$ of the $n + p \to d +\gamma$ reaction, due to a fractional
    change in $x=a_s$ (in red, color online) or $x=B_d$ (in blue)\,:
    $\frac{\partial\log{\gamma}}{\partial\log{x}}$\,. Data points are from
    a numerical evaluation of Eq.~(\ref{eq:crossrate}), the thin solid
    lines represent a rational function fit to these data and the thick
    solid lines represent the powers found in the extreme low energy (or
    low temperature) limit. $T_9 := T / [10^9\,\textnormal{K}]$\,.
  }
\end{figure}

From the formulas above in leading order, \textit{i.e.} at very low
energies, where the $M1$-contribution dominates, the dependence of the
cross section and hence of the rate of the $n + p \to d + \gamma$
reaction is found to be
\begin{equation}
  \gamma_{M1;np\to d\gamma} \propto B_d^{\frac{5}{2}}\,a_s^2\,,
\end{equation}
leading to 
\begin{equation}
  \frac{\partial\log{\gamma}}{\partial\log{a_s}} = 2\,,
  \qquad
  \frac{\partial\log{\gamma}}{\partial\log{B_d}} = \frac{5}{2}\,,
\end{equation}
which are energy or temperature independent.  However, taking into
account the full N${}^3$LO-expressions above, this is found to vary
appreciably in the BBN-relevant range $10^{-3} \le T_9 \le 10$\,,
with  $T_9 := T / [10^9\,\textnormal{K}]$,
leading to a rather strong suppression of the dependence at high
temperatures, see Fig.~\ref{fig:dloggdlogx}\,.

\subsubsection{Temperature dependence of the rates for the leading reactions}

\begin{figure}[!htb] 
  \includegraphics[width=\columnwidth,viewport = 320 470 620 720]{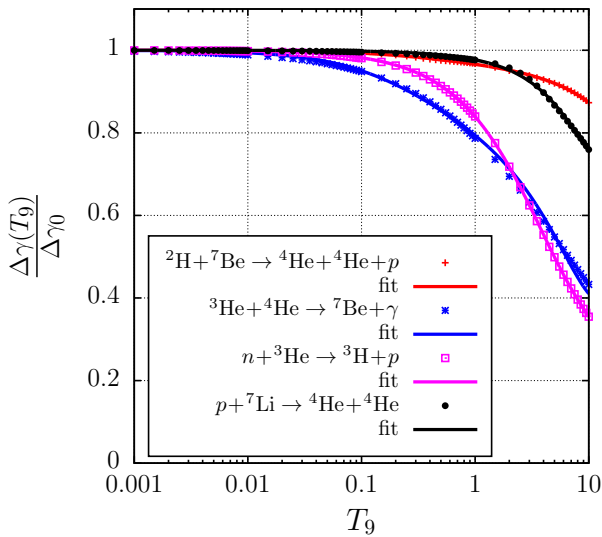}
  \caption{\label{fig:rate_T_dep}%
    Temperature dependence of the change in the rate
    $\Delta\gamma(T_9)/\Delta\gamma_0$ for a fractional change in the
    $Q$-value of some selected leading reactions, due to a fractional
    change in $x=a_s$ or $x=B_d$\,:
    $\frac{\partial\log{\gamma}}{\partial\log{x}}$\,. Data points are from
    a numerical evaluation of Eqs.~(\ref{eq:Qdepx},\ref{eq:radcapx});
    solid lines  (color online) represent the rational function fits
    used in subsequent calculations. $T_9 := T / [10^9\,\textnormal{K}]$\,.
  }
\end{figure}

Such a suppression, in particular for the dependence on the binding
energies, is to be expected for the other nuclear reactions also.
However, for most reactions the data for the rates are given in the
form of parameterizations as a function of $T_9$ or in the form of
tables.  Only if the cross-sections $\sigma(E)$, or, equivalently, the
astrophysical $S(E)$-function is available, it is possible to study the
temperature dependence of a change in the rate $\gamma$ due to a
change in the binding energy or $Q$-value according to
Eqs.~(\ref{eq:Qdepx},\ref{eq:radcapx})\,. To this end we numerically
compute the integrals in these equations and then fit the ratio of the
change in the rate $\Delta\gamma(T_9)/\Delta\gamma_0$ according to
Eqs.~(\ref{eq:Qdepx},\ref{eq:Qdepa}) for direct reactions or
Eqs.~(\ref{eq:radcapx},\ref{eq:radcapa}) for radiative capture
reactions by a rational function of the form
\begin{equation}
  g(T_9)
  =
  \frac{1+r_1\,T_9 + r_2\,T_9^2 + r_3\,T_9^3}{1+q\,T_9}~.
\end{equation}
This is then used in the rate equations. We do this for the leading
reactions in the BBN-network, \textit{viz.} the radiative capture reactions
\begin{eqnarray}
  \label{eq:rcr}
  p+{}^2\textnormal{H}\to{}^3\textnormal{He}+\gamma\,,
  &\quad&
          p+{}^3\textnormal{H}\to{}^4\textnormal{He}+\gamma\,,
  \nonumber\\
  {}^3\textnormal{H}+{}^4\textnormal{He}\to{}^7\textnormal{Li}+\gamma\,,
  &\quad&
          {}^3\textnormal{He}+{}^4\textnormal{He}\to{}^7\textnormal{Be}+\gamma\,,
  \\
  {}^2\textnormal{H}+{}^4\textnormal{He}\to{}^6\textnormal{Li}+\gamma\,,
  &\quad&
          p+{}^7\textnormal{Li}\to{}^4\textnormal{He}+{}^4\textnormal{He}+\gamma\,,
          \nonumber
\end{eqnarray}
the charge exchange reactions
\begin{equation}
  \label{eq:cer}
  n+{}^3\textnormal{He}\to{}^3\textnormal{H}+p\,,
  \quad
  n+{}^7\textnormal{Be}\to{}^7\textnormal{Li}+p\,,
\end{equation}
and the other direct reactions
\begin{eqnarray}
  \label{eq:dr}
  {}^2\textnormal{H}+{}^2\textnormal{H}\to{}^3\textnormal{H}+p\,,
  &\quad&
          {}^2\textnormal{H}+{}^2\textnormal{H}\to{}^3\textnormal{He}+n\,,
  \nonumber \\
  {}^2\textnormal{H}+{}^3\textnormal{H}\to{}^4\textnormal{He}+n\,,
  &\quad&
          {}^2\textnormal{H}+{}^3\textnormal{He}\to{}^4\textnormal{He}+p\,,
  \nonumber\\
  p+{}^7\textnormal{Li}\to{}^4\textnormal{He}+{}^4\textnormal{He}\,,
  &\quad&
          p+{}^6\textnormal{Li}\to{}^3\textnormal{He}+{}^4\textnormal{He}\,,
  \nonumber\\
  {}^2\textnormal{H}+{}^7\textnormal{Be}\to{}^4\textnormal{He}+{}^4\textnormal{He}+p\,,
  &&
\end{eqnarray}
for which parameterizations of the energy dependence of the cross
section can \textit{e.g.} be found in appendix~D of~\cite{Serpico:2004gx}\,.
For some selected reactions the calculated
ratios $\Delta\gamma(T_9)/\Delta\gamma_0$ together with the rational
function fit are shown in Fig.~\ref{fig:rate_T_dep}\,, showing that
for temperatures $T_9 > 0.1$ this change is indeed generally
suppressed with respect to the change calculated with
Eqs.~(\ref{eq:Qdepa},\ref{eq:radcapa})

\section{\label{sec:response} The BBN response matrix}

We estimate the linear dependence of primordial abundances $Y_n$ on
small changes in the neutron life time $\tau_n$, the nuclear binding
energies $B_a$ and the ${}^1 S_0$ $np$ scattering length $a_s$\,,
$\partial\log{Y_n}/\partial\log{X_k}$, by calculating the abundance of the nuclide
\begin{equation}
  n \in \lbrace
  {^2}\textnormal{H}\,,
  {^3}\textnormal{H}+{^3}\textnormal{He}\,,
  {^4}\textnormal{He}\,,
  {^6}\textnormal{Li}\,,
  {^7}\textnormal{Li}+{^7}\textnormal{Be}\,,
  \rbrace
\end{equation}
\textit{i.e.}
\begin{equation}
  Y_n(X_k(1+\delta_k))\,,\qquad
  X_k \in \lbrace \tau_n\,,B_a\,,a_s \rbrace
\end{equation}
for fractional changes $\delta_k \approx \mathcal{O}(10^{-3})$ in the nuclear
parameters $X_k$ with the publicly available codes for BBN: A version
of the Kawano code called \texttt{NUC123}~\cite{Kawano:1992ua} (in
\texttt{FORTRAN}), two more modern implementations based on this:
\texttt{PArthENoPE}~\cite{Pisanti:2007hk,Consiglio:2017pot,Gariazzo:2021iiu} (\texttt{in FORTRAN}) and
\texttt{AlterBBN}~\cite{Arbey:2011nf, Arbey:2018zfh} (in \texttt{C}) as well as an 
implementation as a \texttt{mathematica}-notebook:
\texttt{PRIMAT}~\cite{Pitrou:2018cgg}\,. 
To this end we perform a least-squares fit of a quadratic polynomial to the abundances:
\begin{equation}
  P_k(\delta_k) := c_0\left(1+c_1\,\delta_k + c_2\,\delta_k^2\right)\,,
\end{equation}
such that
\begin{equation}
  \diff{}{c_j}\left|Y_n(X_k(1+\delta_k))-P_k(\delta_k)\right|^2 = 0\,,\quad j=0,1,2\,.
\end{equation}
Then
\begin{equation}
  \diff{\log{Y_n}}{\log{X_k}}
  \approx
  c_1
\end{equation}
will be called an element of the (linear) nuclear BBN response
matrix. It represents the (dimensionless) fractional change in the primordial
abundance $Y_n$ due to a fractional change in the nuclear parameter
$X_k$ in linear approximation.
This method was preferred over a direct approximation of the
(logarithmic) derivatives via finite differences, since some of the
calculated abundances turned out to be rather noisy in particular
when close to zero.

\section{\label{sec:results} Numerical results}

In what follows, we use $\eta=6.14\cdot 10^{-10}$ from~\cite{pdg2022}
as the baryon-to-photon density ratio, but we will also allow for
variations of this parameter. In all programmes, the
following modifications were made:
\begin{itemize}
\item
  All natural constants were updated to recent values listed by
  CODATA~\cite{codata2019} and PDG~\cite{ParticleDataGroup:2020ssz,pdg2022}\,;
\item
  Atomic mass excesses were taken from the
  NUBASE2020~\cite{Huang:2021nwk} compilation; nuclear binding energies
  were then calculated by accounting for the atomic binding energies
  from~\cite{abe2021}\,, although this is a minor effect;
\item
  Reverse reaction rate constants were recalculated on the basis of
  the nuclear data above\,;
\item
  In \texttt{NUC123} hard coded constants were substituted by their
  analytical expressions in terms of natural constants updated from
  CODATA~\cite{codata2019} and
  PDG~\cite{ParticleDataGroup:2020ssz,pdg2022}\,.
\end{itemize}

In addition, and also in the
\texttt{primat}-implementation~\cite{Pitrou:2018cgg}, the rate of the
$n + p \to d + \gamma$ radiative capture reaction was calculated on
the basis of the analytical expression for the cross section from the
EFT calculation in~\cite{Rupak:1999rk}\,, see also section~\ref{sec:bf}
and rational function fits of the temperature dependence and the
dependencies on the deuteron binding energy and the ${}^1 S_0$ $np$
scattering length, as explained in Sect.~\ref{sec:response} were
used. 

For the fractional change in the rate due to changes in the
$Q$-values of direct reactions we used the expression of
Eqs.~(\ref{eq:Qdepa},\ref{eq:radcapa}) with the exception of the
reactions listed in Eqs.~(\ref{eq:rcr},\ref{eq:cer},\ref{eq:dr}),
where we used Eqs.~(\ref{eq:Qdepx},\ref{eq:radcapx})\,. For the
$Q$-value dependence of the weak decay rates we used
Eqs.~(\ref{eq:wdlogf},\ref{eq:wdx})\,.

\begin{table}[hbt]
  \caption{\label{tab:finab}%
    Final abundances as number ratios $Y_n/Y_H$ (for
    ${}^4\textnormal{He}$ the mass ratio $Y_p$) calculated with the modified
    versions of the codes. The nominal reaction $Q_0$-values used were calculated
    from the nuclear binding energies taken from~\cite{Huang:2021nwk}\,,
    furthermore $\eta = 6.14 \cdot 10^{-10}$ and $\tau_n =
    879.4\,\textnormal{s}$\,.
}
\begin{ruledtabular}
\begin{tabular}{lddddd}
  \texttt{code}
  &
    \multicolumn{1}{c}{${}^2\textnormal{H}$}
  &
    \multicolumn{1}{c}{${}^3\textnormal{H}+\!\!{}^3\textnormal{He}$}
  &
    \multicolumn{1}{c}{$Y_p$}
  &
    \multicolumn{1}{c}{${}^6\textnormal{Li}$}
  &
    \multicolumn{1}{c}{${}^7\textnormal{Li}+\!\!{}^7\textnormal{Be}$}
  \\
  &
    \multicolumn{1}{c}{$\times 10^5$}
  &
    \multicolumn{1}{c}{$\times 10^5$}
  &
    \multicolumn{1}{c}{}
  &
    \multicolumn{1}{c}{$\times 10^{14}$}
  &
    \multicolumn{1}{c}{$\times 10^{10}$}
  \\
  \colrule
  \texttt{NUC123}     & 2.550 & 1.040 & 0.247 & 1.101 & 4.577 \\
  \texttt{parthenope} & 2.511 & 1.032 & 0.247 & 1.091 & 4.672 \\
  \texttt{alterbbn}   & 2.445 & 1.031 & 0.247 & 1.078 & 5.425 \\
  \texttt{primat}     & 2.471 & 1.044 & 0.247 & 1.198 & 5.413 \\
  \colrule
  PDG~\cite{pdg2022}  & 2.547 &       & 0.245 &       & 1.6 \\
  $\qquad\pm$  & 0.025 &       & 0.003 &       & 0.3 \\ 
\end{tabular}
\end{ruledtabular}
\end{table}

The values obtained with each of the programmes for the abundances
at the end of the BBN-epoch in terms of the number ratios
$Y_{{}^2\textnormal{H}}/Y_\textnormal{H}$\,,
$Y_{{}^3\textnormal{H}+{}^3\textnormal{He}}/Y_\textnormal{H}$\,,
$Y_{{}^6\textnormal{Li}}/Y_\textnormal{H}$\,,
$Y_{{}^7\textnormal{Li}+{}^7\textnormal{Be}}/Y_\textnormal{H}$\,,
and the mass ratio for ${}^4\textnormal{He}$ are listed in
Tab.~\ref{tab:finab}\,.  The discrepancy of about a factor of three
in the abundance ratio of ${}^7\textnormal{Li}+{}^7\textnormal{Be}$ to
$\textnormal{H}$ is known as the ``$\textnormal{Li}$-problem''\,,
see~\cite{pdg2022} and references therein.

In order to determine the sensitivity of the abundances, each of the
programmes was run for fractional changes of the parameters
$\tau_n$\,, $a_s$\,,
$B_{{}^2\textnormal{H}}$\,,
$B_{{}^3\textnormal{H}}$\,,
$B_{{}^3\textnormal{He}}$\,,
$B_{{}^4\textnormal{He}}$\,,
$B_{{}^6\textnormal{Li}}$\,,
$B_{{}^7\textnormal{Li}}$
and
$B_{{}^7\textnormal{Be}}$
for about a dozen equidistant values in a range $[-\delta,\delta]\,,
\delta\approx \mathcal{O}(10^{-4}-10^{-3})$\,.

The elements of the response matrix were then determined by a
polynomial fit, as explained above in Sect.~\ref{sec:response} for
the abundances of $Y_{{}^2\textnormal{H}}/Y_\textnormal{H}$\,,
$Y_{{}^3\textnormal{H}+{}^3\textnormal{He}}/Y_\textnormal{H}$\,,
$Y_{{}^6\textnormal{Li}}/Y_\textnormal{H}$\,,
$Y_{{}^7\textnormal{Li}+{}^7\textnormal{Be}}/Y_\textnormal{H}$\,,
and the mass ratio for ${}^4\textnormal{He}$\,. 

The values for the resulting response matrix elements are given
and compared to some results from the literature in Tab.~\ref{tab:rme}\,.

\begin{ruledtabular} 
  \begin{longtable}[c]{@{\extracolsep{-0.1em}}llddddd}
    \caption{\label{tab:rme}%
    BBN response matrix $\partial\log{Y_n}/\partial\log{X_k}$ at
    $\eta=6.14 \cdot 10^{-10}$ and $\tau_n =
    879.4\,\textnormal{s}$\,. $Y_n$ are the number ratios of the
    abundances relative to hydrogen; $Y_p$ is conventionally the
    ${}^4$He/H mass ratio. The results obtained with the four BBN codes:
    \texttt{NUC123}\cite{Kawano:1992ua},
    \texttt{alterBBN}\cite{Arbey:2018zfh},
    \texttt{PArthENoPE}\cite{Gariazzo:2021iiu},
    \texttt{PRIMAT}\cite{Pitrou:2018cgg}
    are given in four subsequent rows for each nuclear parameter.
    Below that, for comparison also listed are the values obtained
    in~\cite{Berengut:2013nh} (see Tab.~VII), where $\eta=6.19 \cdot
    10^{-10}$ was used, obtained in ~\cite{Dent:2007zu} (see Tab.~I),
    where $\eta=6.1\cdot 10^{-10}$ and $\tau_n=885.7~\textnormal{s}$ were
    used, as well as the dependence on the deuteron binding energy listed
    in~\cite{Berengut:2009js} (method 2 in Tab.~I), where $\eta=6.22\cdot
    10^{-10}$ was used.  The entries with ${}^{(*)}$ are obtained by
    assuming a $a_s^2$ dependence of the $n + p \to d+ \gamma$ rate, as
    was done in~\cite{Berengut:2013nh}\,.  The entries with
    ${}^{(\dagger)}$ ignore the $T$-dependence of the rates of the leading
    reactions, as also done in~\cite{Berengut:2013nh}\,.
  }\\
  \hline
  \hline
    $X$ 
    &
      \texttt{code}
    &
      \multicolumn{1}{c}{${}^2\textnormal{H}$}
    &
      \multicolumn{1}{c}{${}^3\textnormal{H}\!+\!\!{}^3\textnormal{He}$}
    &
      \multicolumn{1}{c}{${}^4\textnormal{He}$}
    &
      \multicolumn{1}{c}{${}^6\textnormal{Li}$}
    &
      \multicolumn{1}{c}{${}^7\textnormal{Li}\!+\!\!{}^7\textnormal{Be}$}
      \\
      \hline
    \colrule
    \endfirsthead
    \caption{ ... continued ... }\\
    \hline
    $X$
    &
      \texttt{code}
    &
      \multicolumn{1}{c}{${}^2\textnormal{H}$}
    &
      \multicolumn{1}{c}{${}^3\textnormal{H}\!+\!\!{}^3\textnormal{He}$}
    &
      \multicolumn{1}{c}{${}^4\textnormal{He}$}
    &
      \multicolumn{1}{c}{${}^6\textnormal{Li}$}
    &
      \multicolumn{1}{c}{${}^7\textnormal{Li}\!+\!\!{}^7\textnormal{Be}$}
    \\
    \colrule
    \endhead
    \colrule
    \multicolumn{7}{r}{  {\footnotesize ... continued ...} }\\
    \endfoot
    \endlastfoot
    \colrule
    & \texttt{NUC123}     & -0.12 &  0.09 & 0.00 & -0.12 & 0.97 \\
    & \texttt{PArthENoPE} & -0.18 &  0.06 & 0.04 & -0.17 & 1.07 \\
    & \texttt{alterBBN}   & -0.13 &  0.08 & 0.00 & -0.13 & 0.99 \\
    & \texttt{PRIMAT}     & -0.13 &  0.08 & 0.00 & -0.13 & 0.99 \\
    \cline{2-7}
    $a_s$
    & \texttt{NUC123}${}^{(*)}$     & -0.37 &  0.18 &  0.01 & -0.36 & 2.52 \\
    & \texttt{PArthENoPE}${}^{(*)}$ & -0.41 &  0.15 &  0.01 & -0.41 & 2.60 \\
    & \texttt{alterBBN}${}^{(*)}$   & -0.40 &  0.15 &  0.01 & -0.39 & 2.57 \\
    & \texttt{PRIMAT}${}^{(*)}$     & -0.41 &  0.16 &  0.01 & -0.40 & 2.58 \\
    \cline{2-7}
    & Ref.~\cite{Berengut:2013nh}      & -0.39        &  0.17        &  0.01        & -0.38        & 2.64  \\
    \colrule
    & \texttt{NUC123}     & -2.78 & -2.08 & 0.67 & -6.26 & 8.41 \\
    & \texttt{PArthENoPE} & -2.49 & -2.27 & 0.68 & -5.93 & 6.16 \\
    & \texttt{alterBBN}   & -2.93 & -2.09 & 0.69 & -6.38 & 8.76 \\
    & \texttt{PRIMAT}     & -2.89 & -2.11 & 0.69 & -6.33 & 8.63 \\
    \cline{2-7}
    $B_{{}^2\!\textnormal{H}}$
    & \texttt{NUC123}${}^{(\dagger)}$     & -2.74 & -2.08 &  0.67 & -6.41 & 8.36 \\
    & \texttt{PArthENoPE}${}^{(\dagger)}$ & -2.45 & -2.28 &  0.68 & -6.10 & 6.11 \\
    & \texttt{alterBBN}${}^{(\dagger)}$   & -2.89 & -2.10 &  0.68 & -6.54 & 8.71 \\
    & \texttt{PRIMAT}${}^{(\dagger)}$     & -2.85 & -2.12 &  0.68 & -6.49 & 8.59 \\
    \cline{2-7}
    & Ref.~\cite{Berengut:2013nh}      & -2.91        &  -2.08        &  0.67        & -6.57        & 9.44  \\
    & Ref.~\cite{Dent:2007zu}          & -2.8         &  -2.1         &  0.68        & -6.8         & 8.8  \\
    & Ref.~\cite{Berengut:2009js}      & -2.91        &  -2.1         &  0.67        & -6.58        & 9.41  \\
    \colrule
    & \texttt{NUC123}     & -0.32 & -2.22 &  0.01 & -0.31 & -3.61 \\
    & \texttt{PArthENoPE} & -0.28 & -2.14 & -0.02 & -0.27 & -3.49 \\
    & \texttt{alterBBN}   & -0.26 & -2.16 &  0.00 & -0.25 & -3.60 \\
    & \texttt{PRIMAT}     & -0.28 & -2.20 &  0.01 & -0.27 & -3.58 \\
    \cline{2-7}
    $B_{{}^3\!\textnormal{H}}$
    & \texttt{NUC123}${}^{(\dagger)}$     & -0.31 & -2.39 &  0.01 & -0.29 & -3.93 \\
    & \texttt{PArthENoPE}${}^{(\dagger)}$ & -0.27 & -2.29 &  0.03 & -0.26 & -3.80 \\
    & \texttt{alterBBN}${}^{(\dagger)}$   & -0.25 & -2.32 &  0.01 & -0.24 & -3.92 \\
    & \texttt{PRIMAT}${}^{(\dagger)}$     & -0.27 & -2.37 &  0.01 & -0.25 & -3.90 \\
    \cline{2-7}
    & Ref.~\cite{Berengut:2013nh}      & -0.27        &  -2.36        &  0.01        & -0.26        & -3.84  \\
    & Ref.~\cite{Dent:2007zu}          & -0.22        &  -1.4         &  0           & -0.20        & -2.5  \\
    \colrule
    & \texttt{NUC123}     & -2.35 &  3.76 &  0.01 & -5.49 & -6.47 \\
    & \texttt{PArthENoPE} & -2.37 &  3.67 &  0.04 & -5.61 & -6.57 \\
    & \texttt{alterBBN}   & -2.42 &  3.68 &  0.01 & -5.66 & -6.72 \\
    & \texttt{PRIMAT}     & -2.36 &  3.73 &  0.01 & -5.61 & -6.75 \\
    \cline{2-7}
    $B_{{}^3\!\textnormal{He}}$
    & \texttt{NUC123}${}^{(\dagger)}$     & -2.44 &  3.94 &  0.01 & -5.66 & -8.13 \\
    & \texttt{PArthENoPE}${}^{(\dagger)}$ & -2.73 &  3.76 &  0.11 & -6.16 & -8.06 \\
    & \texttt{alterBBN}${}^{(\dagger)}$   & -2.58 &  3.91 &  0.01 & -5.91 & -8.37 \\
    & \texttt{PRIMAT}${}^{(\dagger)}$     & -2.46 &  3.91 &  0.01 & -5.79 & -8.41 \\
    \cline{2-7}
    & Ref.~\cite{Berengut:2013nh}      & -2.38        &  3.85        &  0.01        & -5.72        & -8.27  \\
    & Ref.~\cite{Dent:2007zu}          & -2.1         &  3.0         &  0           & -3.1         & -9.5   \\
    \colrule
    & \texttt{NUC123}     & -0.01 & -0.79 &  -0.00 & -66.73 & -49.91 \\
    & \texttt{PArthENoPE} & -0.02 & -0.83 &   0.00 & -66.67 & -50.14 \\
    & \texttt{alterBBN}   & -0.08 & -0.85 &   0.00 & -66.72 & -50.90 \\
    & \texttt{PRIMAT}     & -0.02 & -0.82 &   0.00 & -66.91 & -50.48 \\
    \cline{2-7}
    $B_{{}^4\!\textnormal{He}}$
    & \texttt{NUC123}${}^{(\dagger)}$     & -0.01 & -0.80 & -0.00 & -69.49 & -57.39 \\
    & \texttt{PArthENoPE}${}^{(\dagger)}$ & -0.02 & -0.84 &  0.01 & -69.53 & -57.64 \\
    & \texttt{alterBBN}${}^{(\dagger)}$   & -0.09 & -0.86 &  0.00 & -69.49 & -58.43 \\
    & \texttt{PRIMAT}${}^{(\dagger)}$     & -0.02 & -0.83 &  0.00 & -69.70 & -58.06 \\
    \cline{2-7}
    & Ref.~\cite{Berengut:2013nh}      & -0.03        &  -0.84        &  0.00        & -69.8        & -57.4  \\
    & Ref.~\cite{Dent:2007zu}          & -0.01         &  -0.57       &  0           & -59          & -57    \\
    \colrule
    & \texttt{NUC123}     & -0.00 &  0.00 &   0.00 &  75.40 & 0.00 \\
    & \texttt{PArthENoPE} & -0.01 & -0.01 &   0.01 &  75.34 & 0.02 \\
    & \texttt{alterBBN}   &  0.00 & -0.01 &   0.00 &  75.35 & 0.01 \\
    & \texttt{PRIMAT}     & -0.00 &  0.00 &   0.00 &  75.65 & 0.00 \\
    \cline{2-7}
    $B_{{}^6\!\textnormal{Li}}$
    & \texttt{NUC123}${}^{(\dagger)}$     & -0.00 &  0.00 &   0.00 &  78.51 &  0.00 \\
    & \texttt{PArthENoPE}${}^{(\dagger)}$ &  0.02 &  0.01 &  -0.01 &  78.60 & -0.02 \\
    & \texttt{alterBBN}${}^{(\dagger)}$   &  0.00 & -0.00 &   0.00 &  78.48 &  0.00 \\
    & \texttt{PRIMAT}${}^{(\dagger)}$     & -0.00 &  0.00 &   0.00 &  78.81 &  0.00 \\
    \cline{2-7}
    & Ref.~\cite{Berengut:2013nh}      &  0.00        &  0.00        &  0.00        & 78.9        & 0.00  \\
    & Ref.~\cite{Dent:2007zu}          &  0           &  0           &  0           & 69          & 0  \\
    \colrule
    & \texttt{NUC123}     & -0.00 & -0.00 &  -0.00 & -0.00 & -22.65 \\
    & \texttt{PArthENoPE} & -0.02 & -0.01 &   0.02 & -0.02 & -23.12 \\
    & \texttt{alterBBN}   &  0.01 & -0.00 &  -0.00 &  0.01 & -23.17 \\
    & \texttt{PRIMAT}     & -0.00 & -0.00 &   0.00 & -0.00 & -23.39 \\
    \cline{2-7}
    $B_{{}^7\!\textnormal{Li}}$
    & \texttt{NUC123}${}^{(\dagger)}$     & -0.00 & -0.00 &  -0.00 &  -0.00 & -23.54 \\
    & \texttt{PArthENoPE}${}^{(\dagger)}$ &  0.01 &  0.01 &  -0.01 &   0.01 & -24.09 \\
    & \texttt{alterBBN}${}^{(\dagger)}$   &  0.01 &  0.00 &  -0.00 &   0.01 & -24.09 \\
    & \texttt{PRIMAT}${}^{(\dagger)}$     & -0.00 &  0.00 &   0.00 &  -0.00 & -24.31 \\
    \cline{2-7}
    & Ref.~\cite{Berengut:2013nh}      &  0.03        &  0.01        &  0.00        & 0.02        & -25.1  \\
    & Ref.~\cite{Dent:2007zu}          &  0           &  0           &  0            & 0          & -6.9  \\
    \colrule
    & \texttt{NUC123}     & -0.00 & -0.00 &   0.00 & -0.00 &  86.58 \\
    & \texttt{PArthENoPE} &  0.02 &  0.00 &  -0.01 &  0.01 &  87.31 \\
    & \texttt{alterBBN}   &  0.00 &  0.00 &  -0.00 &  0.00 &  88.47 \\
    & \texttt{PRIMAT}     & -0.00 & -0.00 &  -0.00 &  0.00 &  88.38 \\
    \cline{2-7}
    $B_{{}^7\!\textnormal{Be}}$
    & \texttt{NUC123}${}^{(\dagger)}$     & -0.00 & -0.00 &   0.00 &  -0.00 &  97.43 \\
    & \texttt{PArthENoPE}${}^{(\dagger)}$ & -0.02 & -0.01 &   0.00 &  -0.03 &  98.25 \\
    & \texttt{alterBBN}${}^{(\dagger)}$   &  0.00 & -0.00 &  -0.00 &   0.00 &  99.43 \\
    & \texttt{PRIMAT}${}^{(\dagger)}$     &  0.00 & -0.00 &  -0.00 &   0.00 &  99.33 \\
    \cline{2-7}
    & Ref.~\cite{Berengut:2013nh}      &  0.00        &  0.00        &  0.00        & 0.00        & 99.1  \\
    & Ref.~\cite{Dent:2007zu}          &  0           &  0           &  0            & 0          & 81  \\
    \colrule
    & \texttt{NUC123}     &  0.41 &  0.14 &   0.72 &  1.36 &  0.44 \\
    & \texttt{PArthENoPE} &  0.41 &  0.14 &   0.74 &  1.37 &  0.45 \\
    $\tau_n$
    & \texttt{alterBBN}   &  0.42 &  0.14 &   0.73 &  1.38 &  0.43 \\
    & \texttt{PRIMAT}     &  0.42 &  0.14 &   0.73 &  1.38 &  0.44 \\
    \colrule
    & Ref.~\cite{Berengut:2013nh}      &  0.41        &  0.14        &  0.72        & 1.36        & 0.43  \\
    & Ref.~\cite{Dent:2007zu}          &  0.41        &  0.15        &  0.73        & 1.4         & 0.43  \\
    \colrule
    \hline
  \end{longtable} 
\end{ruledtabular}

\begin{figure*}[htp] 
  \includegraphics[width=1.0\linewidth,viewport = 25 200 575 725]{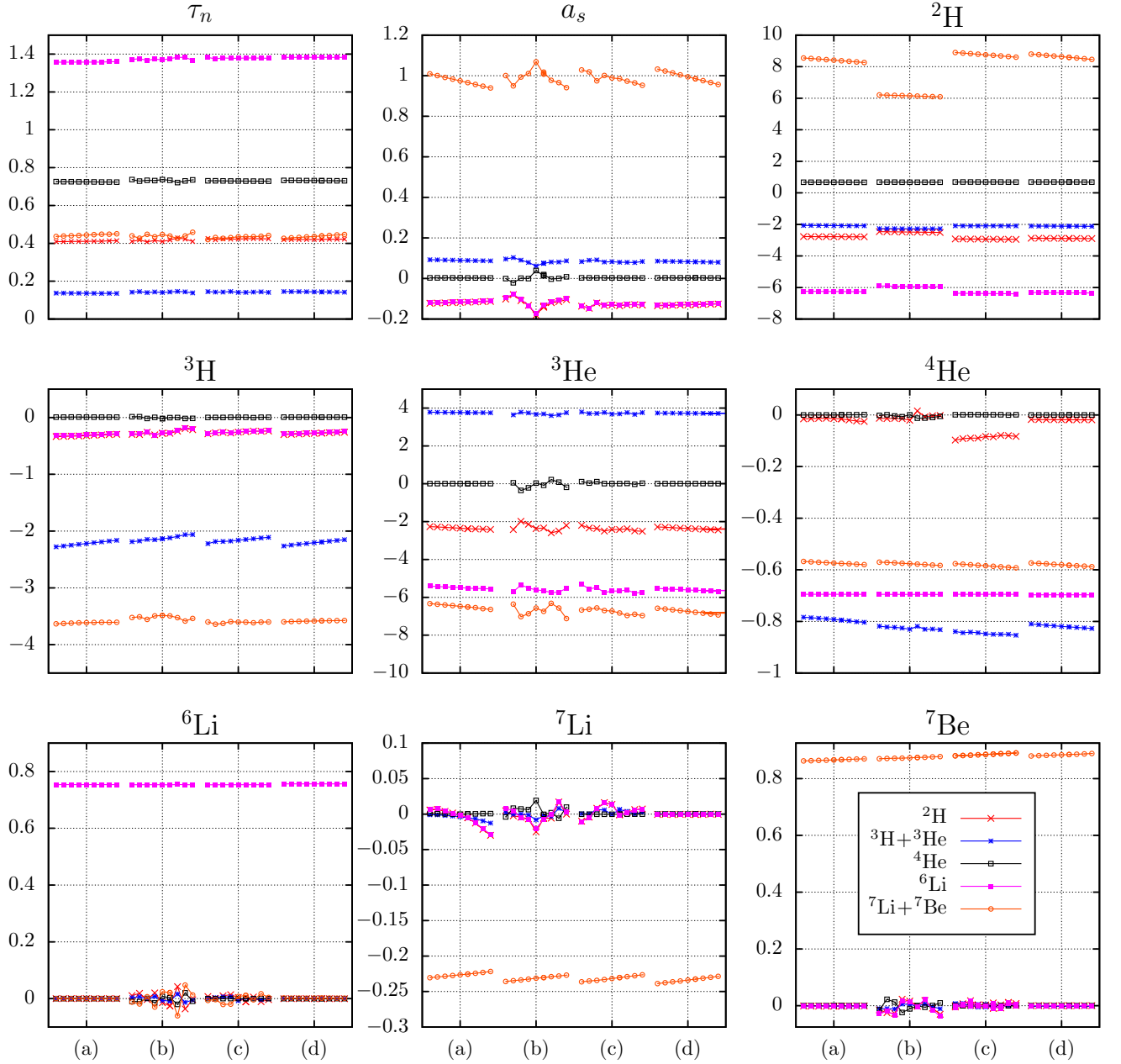}
  \caption{\label{fig:etadep_rme}%
    Baryon-to-photon density $\eta$-dependence of the transfer matrix
    elements $\partial\log{Y_n}/\partial\log{X_k}$\,,
    $n={}^2\textnormal{H}$ (crosses in red),
    ${}^3\textnormal{H}+{}^3\textnormal{He}$ (asterisks in blue),
    ${}^4\textnormal{He}$ (open squares in black),
    ${}^6\textnormal{Li}$ (filled squares in magenta) and
    ${}^7\textnormal{Li}+{}^7\textnormal{Be}$ (open circles in orange; color online)\,.
        For each of the programmes:
    (a) \texttt{NUC123},
    (b) \texttt{PArtHEnoPE},
    (c) \texttt{alterBBN},
    (d) \texttt{PRIMAT},
    the variation of the response matrix elements is shown as 9
    subsequent points for
    $\eta_{10} := \eta\cdot 10^{10} = 5.94, 5.99, 6.04, 6.09, 6.14, 6.19, 6.24, 6.29, 6.34$\,,
    the central values at 6.14 corresponding to the values listed in Tab.~\ref{tab:rme}\,.
    For better visibility, in the panels labeled
    ``${}^4\textnormal{He}$'' and ``${}^6\textnormal{Li}$'' the values of
    the response matrix elements
    $\partial\log{Y_{B_{{}^6\textnormal{Li}}}}/\partial\log{X_k}$ (filled
    squares, in magenta) have been divided by $100$\,. The same applies to
    the panels labeled ``${}^4\textnormal{He}$''\,,
    ``${}^7\textnormal{Li}$'' and ``${}^7\textnormal{Be}$'' for the
    response matrix elements
    $\partial\log{Y_{{}^7\textnormal{Li}+{}^7\textnormal{Be}}}/\partial\log{X_k}$
    (open circles, in orange)\,.
  }
\end{figure*}

In order to illustrate the dependence on the baryon-to-photon density
parameter $\eta$ we displayed in Fig.~\ref{fig:etadep_rme} the
variation of the response matrix elements for a change of $\eta$ in
the range $5.94\cdot10^{-10}-6.34\cdot10^{-10}$ which corresponds to
the error range quoted in~\cite{pdg2022}\,:
$\eta_{10} := \eta\cdot 10^{10} = 6.143 \pm 0.190$\,.

The following observations can be made:
\begin{itemize}
\item
  Concerning the baryon-to-photon density ratio $\eta$-dependence\,:
  Although the final abundances at the end of the primordial
  nucleosynthesis do depend on $\eta$ strong enough and, in particular
  the ${}^4\textnormal{He}$ and deuteron abundances allow for estimating
  a rather strict bound on $\eta$, see \textit{e.g.}  Fig.~24.1 in
  section 24 of~\cite{pdg2022} compatible with the CMB determination of
  the cosmic baryon density~\cite{pdg2022}\,, the $\eta$-dependence of
  the linear response to the nuclear parameters studied here was found
  to be in general almost linear and of rather moderate size, see
  Fig.~\ref{fig:etadep_rme}\,. Note that a conspicuous correlation
  between the values for the response matrix elements exists (many
  values being even almost identical, see Fig.~\ref{fig:etadep_rme})
  for the ${}^2\textnormal{H}$ and ${}^6\textnormal{Li}$ abundances
  for all dependences studied here.
\item
  The values found for the current best value $\eta=6.14\cdot
  10^{-10}$~\cite{pdg2022} were found to be by and large consistent with
  those cited in~\cite{Berengut:2013nh}, if the same assumptions
  concerning the $a_s$-dependence (values indicated by ${}^{(*)}$ in
  Tab.~\ref{tab:rme}) and the temperature independent modification of
  the rates according to Eqs.~(\ref{eq:Qdepa},\ref{eq:radcapa}) are made
  (extracted from the entries indicated by ${}^{\dagger}$ in
  Tab.~\ref{tab:rme})\,, see Tab.~\ref{tab:rmecomp}\,. Note that
  in~\cite{Berengut:2013nh} a slightly different value $\eta=6.19\cdot
  10^ {-10}$ was used.
  
\begin{table}[hbt]
  \caption{\label{tab:rmecomp}%
    Comparison of the range of values obtained with the four codes
    studied here with the results of~\cite{Berengut:2013nh}\,. For each
    parameter $X$ the first two lines then give the minimal and maximal
    values obtained with the four codes, on the basis of the same
    assumptions as in~\cite{Berengut:2013nh}\, \textit{i.e.} these were 
    extracted from the entries labeled ${}^{(*)}$ and ${}^{(\dagger)} $ in
    Tab.~\ref{tab:rme}\,. The third line below the
    horizontal rule contains the values reported in~\cite{Berengut:2013nh}\,.
  }
  \begin{ruledtabular}
    \begin{tabular}{lddddd}
      \texttt{$X$}
      &
        \multicolumn{1}{c}{${}^2\textnormal{H}$}
      &
        \multicolumn{1}{c}{${}^3\textnormal{H}+\!\!{}^3\textnormal{He}$}
      &
        \multicolumn{1}{c}{$Y_p$}
      &
        \multicolumn{1}{c}{${}^6\textnormal{Li}$}
      &
        \multicolumn{1}{c}{${}^7\textnormal{Li}+\!\!{}^7\textnormal{Be}$}
      \\
      \colrule
                           & -0.41 & 0.15 & 0.01 & -0.41 & 2.52 \\
      $a_s$                & -0.37 & 0.18 & 0.01 & -0.36 & 2.60 \\
      \cline{2-6}
                           & -0.39 & 0.17 & 0.01 & -0.38  & 2.64 \\
      \colrule
                           & -2.89 & -2.28 & 0.67 & -6.54 & 6.11 \\
      $B_{{}^2\!\textnormal{H}}$ & -2.45 & -2.08 & 0.68 & -6.10 & 8.71 \\
      \cline{2-6}
                           & -2.91 & -2.08 & 0.67 & -6.57 & 9.44 \\
      \colrule  
                           & -0.31 & -2.39 & 0.01 & -0.29 & -3.93 \\ 
      $B_{{}^3\!\textnormal{H}}$ & -0.25 & -2.29 & 0.03 & -0.24 & -3.80 \\
      \cline{2-6}
                           & -0.27 & -2.36 & 0.01 & -0.26 & -3.84 \\
      \colrule  
                           & -2.73 & 3.76 & 0.01 & -6.16 & -8.41 \\ 
      $B_{{}^3\!\textnormal{He}}$ & -2.44 & 3.94 & 0.11 & -5.66 & -8.06 \\
      \cline{2-6}
                           & -2.38 & 3.85  & 0.01 & -5.72 & -8.27 \\
      \colrule  
                           & -0.09 & -0.86 & -0.00 & -69.7 & -58.4 \\ 
      $B_{{}^4\!\textnormal{He}}$ & -0.01 & -0.80 &  0.01 & -69.5 & -57.4 \\
      \cline{2-6}
                           & -0.03 &  -0.84 & 0.00 & -69.8 & -57.4 \\
      \colrule  
                           & -0.00 & -0.00 & -0.01 & 78.5 & -0.02 \\ 
      $B_{{}^6\!\textnormal{Li}}$ &  0.02 &  0.01 &  0.00 & 78.8 &  0.00 \\
      \cline{2-6}
                           & 0.00  & 0.00  & 0.00 &  78.9  & 0.00 \\
      \colrule  
                           & -0.00 & -0.00 & -0.01 & -0.00  & -24.3 \\ 
      $B_{{}^7\!\textnormal{Li}}$ &  0.01 &  0.01 &  0.00 & 0.01   & -23.5 \\
      \cline{2-6}
                           & 0.03 & 0.01 & 0.00 & 0.02 & -25.1 \\ 
      \colrule
                           & -0.02 & -0.01 & -0.00 & -0.03  & 97.4 \\ 
      $B_{{}^7\!\textnormal{Be}}$ &  0.00 & -0.00 &  0.00 & 0.00 & 99.4 \\
      \cline{2-6}
                           &   0.00 & 0.00 & 0.00 &   0.00  & 99.1 \\
      \colrule  
                          & 0.41 &   0.14 & 0.72 & 1.36 & 0.43 \\ 
      $\tau_n$            & 0.42 &   0.14 & 0.74 & 1.38 & 0.45 \\
      \cline{2-6}
                          & 0.41 &   0.14 & 0.72 & 1.36 & 0.43 \\
    \end{tabular}
  \end{ruledtabular}
\end{table}
\item
  More importantly, it was found that accounting for the actual $a_s$
  dependence in the calculation of the cross section for the $n + p \to
  d + \gamma$ reaction and, accordingly for a temperature dependence in
  the corresponding rate, in fact the linear dependence of the
  abundances on this scattering length is reduced with respect to the
  temperature independent $a_s^2$ dependence assumed
  in~\cite{Berengut:2013nh} approximately by a factor of three.
\item
  Also, accounting for the temperature dependence in the rate when
  varying the bindings energies of at least the leading reactions in the
  nuclear BBN network, reduces the values of the response matrix
  elements in some cases appreciably: In particular this applies to the
  abundance-dependence of ${}^6\textnormal{Li}$ in the binding energy
  of ${}^6\textnormal{Li}$ ($\approx 80 \mapsto \approx 75$) and the
  abundance-dependences of ${}^7\textnormal{Li}+{}^7\textnormal{Be}$ on
  the binding energy of ${}^3\textnormal{H}$ ($\approx 4 \mapsto \approx
  3.5$), on the binding energy of ${}^3\textnormal{He}$ ($\approx 8.25
  \mapsto \approx 6.75$) and on the binding energy of
  ${}^7\textnormal{Be}$ ($\approx 100 \mapsto 87$)\,.
\item
  Furthermore, although the numbers found are roughly consistent, see
  \textit{e.g.} the ranges listed in
  Tabs.~\ref{tab:rmecomp},\ref{tab:rmerng}\,, the four publicly
  available codes, which do differ in the number of coupled reactions
  treated in the BBN-network, the treatment of radiative corrections to
  processes as well as the specific parameterization of the nuclear
  reaction rates, do lead to slightly different values for the response
  matrix elements studied here.
  
  \begin{table}[hbt]
    \caption{\label{tab:rmerng}%
      Model dependence of the response matrix elements. Listed are the
      ranges for results extracted from Tab.~\ref{tab:rme}\,.  For each
      parameter $X$ the upper row gives the minimal and the lower row the
      maximal value found with any of the four codes.
    }
    \begin{ruledtabular}
      \begin{tabular}{lddddd}
        \texttt{$X$}
        &
          \multicolumn{1}{c}{${}^2\textnormal{H}$}
        &
          \multicolumn{1}{c}{${}^3\textnormal{H}+\!\!{}^3\textnormal{He}$}
        &
          \multicolumn{1}{c}{$Y_p$}
        &
          \multicolumn{1}{c}{${}^6\textnormal{Li}$}
        &
          \multicolumn{1}{c}{${}^7\textnormal{Li}+\!{}^7\textnormal{Be}$}
        \\
        \colrule
        $a_s$                & -0.18 & 0.06 & 0.00 & -0.17 & 0.97 \\
                             & -0.12 & 0.09 & 0.04 & -0.12 & 1.07 \\
        \colrule
        $B_{{}^2\!\textnormal{H}}$ & -2.93 & -2.27 & 0.67 & -6.38 & 6.16 \\
                             & -2.49 & -2.07 & 0.69 & -5.93 & 8.76 \\
        \colrule  
        $B_{{}^3\!\textnormal{H}}$ & -0.32 & -2.22 & -0.02 & -0.31 & -3.61 \\ 
                            & -0.26 & -2.14 &  0.01 & -0.25 & -3.49 \\
        \colrule  
        $B_{{}^3\!\textnormal{He}}$ & -2.42 & 3.67 & 0.01 & -5.66 & -6.75 \\ 
                             & -2.35 & 3.76 & 0.04 & -5.49 & -6.47 \\
        \colrule  
        $B_{{}^4\!\textnormal{He}}$ & -0.08 & -0.85 & -0.00 & -66.9 & -50.9 \\ 
                              & -0.01 & -0.79 &  0.00 & -66.7 & -49.9 \\
        \colrule  
        $B_{{}^6\!\textnormal{Li}}$ &  -0.01 & -0.01 & 0.00 & 75.3 & 0.00 \\ 
                              &  0.00 & 0.00 & 0.01 & 75.7 &  0.02 \\
        \colrule  
        $B_{{}^7\!\textnormal{Li}}$ & -0.02 & -0.01 & -0.00 & -0.02  & -23.4 \\ 
                              & 0.01 & -0.00 & 0.02 & 0.01 & -22.7 \\
        \colrule  
        $B_{{}^7\!\textnormal{Be}}$  & -0.00 & -0.00 & -0.01 & -0.00 & 86.6 \\ 
                               & 0.02 &   0.00 & 0.00    & 0.01 & 88.5 \\ 
        \colrule  
        $\tau_n$              & 0.41 &   0.14 & 0.72 & 1.36 & 0.43 \\ 
                              & 0.42 &   0.14 & 0.74 & 1.38 & 0.45 \\ 
      \end{tabular}
    \end{ruledtabular}
  \end{table}
  
  We shall therefore consider these differences, see
  Tab.~\ref{tab:rmerng}, as an estimate of the systematic (in fact
  model-dependent) deviations for the numbers obtained. This is one of
  the main results of the present investigation. As can also be seen
  from Fig.~\ref{fig:etadep_rme} the largest relative deviations are
  found for the dependence of the
  ${}^7\textnormal{Li}+{}^7\textnormal{Be}$ abundance on the binding
  energy of ${}^2\textnormal{H}$ with the programme \texttt{PArtHEnoPE}:
  20 \%\,. No obvious explanation for this could be found. 
  Also the dependence of the ${}^2\textnormal{H}$ abundance on the
  binding energy of ${}^4\textnormal{He}$ with the programme
  \texttt{alterBBN} deviates from those found with the other
  codes. Note, however, that here an exceptional non-linear dependence on
  $B_{{}^4\textnormal{He}}$ was found; determining the logarithmic
  derivative only from left-sided finite differences yields an almost vanishing
  value, in accordance with the values found with the other programmes.
  The dependence of the ${}^3\textnormal{H}+{}^3\textnormal{He}$
  abundance on the binding energy of ${}^4\textnormal{He}$ varies
  between all programmes by $\approx 8 \%$\,.
\end{itemize}

\section{\label{sec:summary}Summary and conclusion}

We reexamined the response matrix elements $\partial\log{Y_n}
/ \partial\log{X_k}$\,, \textit{i.e.} the linear fractional change in
the abundance $Y_n$ of the nuclides
${}^2\textnormal{H}$\,,
${}^3\textnormal{H} + {}^3\textnormal{He}$\,,
${}^4\textnormal{He}$\,,
${}^6\textnormal{Li}$ and
${}^7\textnormal{Li}+{}^7\textnormal{Be}$
due to a fractional change in nuclear parameters $X_k$\,: the life
time of the neutron $\tau_n$\,, the ${}^1 S_0$ $np$ scattering length
$a_s$ and the binding energies of
${}^2\textnormal{H}$\,,
${}^3\textnormal{H}$\,,
${}^3\textnormal{He}$\,,
${}^4\textnormal{He}$\,,
${}^6\textnormal{Li}$\,,
${}^7\textnormal{Li}$ and 
${}^7\textnormal{Be}$\,.
In addition the dependence of these quantities on the baryon-to-photon
density ratio $\eta$ was studied. In order to obtain an estimate for
the model dependence of the response matrix elements, these were
determined with four publicly available codes for calculating the
abundances of light elements in primordial nucleosynthesis.  The
calculated values were found to be by and large mutually consistent, the
largest deviations were found for matrix elements that almost vanish
anyway. Overall systematic deviations between the codes of a few
percent do occur, however.

In the present treatment the nominal values of nuclear binding
energies (and hence of nominal $Q_0$-values) were updated according to
the most recent nuclear data bases currently available. Moreover, in
contrast to previous studies, we did account for temperature
dependences of the sensitivity of the rate of the leading $n + p \to
d + \gamma$ reaction on the ${}^1 S_0$ $np$ scattering length $a_s$\,,
as well as that of the sensitivity of the rates of a dozen other
leading reactions in the BBN-network to the $Q$-values of these
reactions. Both effects lead to a reduction of the magnitude of the
response matrix elements: the first effect to a reduction by a factor
of three, the latter in some cases approximately by 10\%\,. The $\eta$
dependence of the response matrix elements was found to be a minor
effect only.

These findings should be taken into account before making
\textit{e.g.} claims on bounds from primordial nucleosynthesis on more
fundamental parameters, such as quark masses or coupling constants,
underlying the nuclear parameters studied here. Here we shall refrain
from determining such bounds, postponing that to a future publication
and merely mention an important issue to be addressed prior to that.
Ideally, a study of varying nuclear parameters in order to investigate
their impact on abundances in BBN-nucleosynthesis should rely on
accurate theoretical descriptions of at least the major reactions,
such that the dependence of the cross sections of the leading
reactions on nuclear parameters such as scattering lengths, effective
range parameters and binding energies can be determined in
detail. Presently such a description is available for the leading $n+p
\to d+ \gamma$ reaction only in the form of an accurate
EFT-treatment. Therefore this was elaborated on in some detail here.
A similar treatment of other reactions is the subject of current
research, see \textit{e.g.}  the discussion of $\alpha-\alpha$
scattering in~\cite{Elhatisari:2021eyg}\,. A first application
of the insight obtained here will be a reassessment of the
dependence of primordial abundances on the electromagnetic fine-structure
constant~\cite{MMM}.

\begin{acknowledgments}
This project is part of the ERC Advanced Grant ``EXOTIC'' supported
the European Research Council (ERC) under the European Union's Horizon
2020 research and innovation programme (grant agreement
No. 101018170), We further acknowledge support by the Deutsche
Forschungsgemeinschaft (DFG, German Research Foundation) and the NSFC
through the funds provided to the Sino-German Collaborative Research
Center TRR110 ``Symmetries and the Emergence of Structure in QCD''
(DFG Project ID 196253076 - TRR 110, NSFC Grant No. 12070131001), the
Chinese Academy of Sciences (CAS) President's International Fellowship
Initiative (PIFI) (Grant No. 2018DM0034) and Volkswagen Stiftung
(Grant No. 93562).
 \end{acknowledgments}

\end{document}